\documentclass[lettersize,journal]{IEEEtran}
\usepackage{amsmath,amsfonts}
\usepackage{array}
\usepackage{cite}
\usepackage[caption=false,font=normalsize,labelfont=sf,textfont=sf]{subfig}
\usepackage{textcomp}
\usepackage{stfloats}
\usepackage{url}
\usepackage{xcolor}
\usepackage{verbatim}
\usepackage{graphicx}
\usepackage{enumitem}
\usepackage{tcolorbox}
\usepackage[justification=centering]{caption}
\usepackage[colorlinks=true,linkcolor=blue,citecolor=blue,urlcolor=black,]{hyperref}
\usepackage{algorithm}
\usepackage{algorithmic}
\usepackage{booktabs}
\usepackage{tabularx}
\usepackage{pifont}
\usepackage[normalem]{ulem}
\usepackage{threeparttable}
\useunder{\uline}{\ul}{}
\renewcommand{\arraystretch}{1.5}

\begin{document}

\title{Towards Personal Data Sharing Autonomy: \linebreak
A Task-driven Data Capsule Sharing System}

\author{Qiuyun Lyu, Yilong Zhou, Yizhi Ren, Zheng Wang, and Yunchuan Guo
\thanks{This work was supported by the Natural Science Foundation of Zhejiang Province (No. LY23F020017) 
and the ``Pioneer'' and ``Leading Goose'' R\&D Program of Zhejiang (Grant No.  2022C03174).}
}

\markboth{Journal of \LaTeX\ Class Files,~Vol.~16, No.~?, August~2024}%
{Shell \MakeLowercase{\textit{et al.}}: A Sample Article Using IEEEtran.cls for IEEE Journals}


\maketitle

\begin{abstract} 
  Personal data custodian services enable data owners to share their data with 
  data consumers in a convenient manner, anytime and anywhere. 
  However, with data hosted in these services being beyond the control of the data 
  owners, it raises significant concerns about privacy in personal data sharing. 
  Many schemes have been proposed to realize fine-grained access control and 
  privacy protection in data sharing. However, they fail to 
  protect the rights of data owners to their data under the law, 
  since their designs focus on the 
  management of system administrators rather than enhancing the data owners' privacy. 
  In this paper, we introduce a novel task-driven 
  personal data sharing system based on the data capsule paradigm realizing personal data sharing 
  autonomy. It enables data owners in our system to fully control their data, and 
  share it autonomously. 
  Specifically, we present a tamper-resistant data capsule encapsulation method, 
  where the data capsule is the minimal unit for independent and secure personal data storage 
  and sharing. 
  Additionally, to realize selective sharing and informed-consent based 
  authorization, we propose a task-driven data sharing mechanism that is resistant to 
  collusion and EDoS attacks. 
  Furthermore, by updating parts of the data capsules, 
  the permissions granted to data consumers can be immediately revoked. 
  Finally, we conduct a security and performance analysis, proving that 
  our scheme is correct, sound, and secure, as well as revealing 
  more advantageous features in practicality, compared with the state-of-the-art schemes. 
\end{abstract}


\section{Introduction}
\IEEEPARstart{P}{ersonal} data, defined as ``any information relating to an 
identified or identifiable natural person"\cite{world2014rethinking}, 
is gathered and maintained by numerous services, 
including online social networks, 
shopping platforms, 
consumer electronics\cite{li2024efficient}, and VANETs\cite{yan2023edge},\cite{wang2023lightweight}. 
To efficiently manage and store this personal data, 
cloud-based Personal Data Storage (PDS\cite{fallatah2023personal}) systems (e.g., Mydex\cite{mydex2010case}, 
WebBox\cite{van2012decentralized}, and HAT\cite{hat2015hat}) 
have been developed. 
These systems enable personal data owners to store their data and 
share it with service providers in exchange for various 
services\cite{fernandez2022privacy}. \par
However, according to the ``cloud attacks rise but most sensitive data remains 
unencrypted'' by Techcircle\cite{techfunnel}, only about 45\% of 
sensitive data hosted in the cloud are encrypted. And the prevalence of data breaches, 
insider attacks, and the misuse of collected data by organizations also show 
that data privacy has emerged as a fundamental challenge in today's digital landscape
\cite{wang2019data}. 
As a result, personal data owners are increasingly 
concerned about losing control of their data. 
This concern is rooted in the fact that data hosted in the cloud 
is physically uncontrollable by owners\cite{xu2023adaptively}. \par
To address the above challenges and ease the concern of data owners, 
governments and organizations have formulated many data privacy regulations, 
such as the General Data Protection Regulation (GDPR\cite{voigt2017eu}), 
the Health Insurance Portability and Accountability Act (HIPAA\cite{ness2007influence}), and 
the California Consumer Privacy Act (CCPA\cite{goldman2020introduction}), 
to protect personal data owners' right to their data. 
From the regulations, we can outline three requirements 
for data privacy protection in the process of personal data sharing, 
namely 1) data collecting and processing minimization, 
which requires data consumers only collect and process the minimum 
data strictly needed for the specified purposes; 2) explicit consent, 
requiring data consumers to obtain unambiguous consent from data owners before accessing data; 
and 3) the right to be forgotten, which necessitates data consumers delete the personal data 
after services are finished. 
These requirements can be concluded into personal data sharing autonomy which 
balances the fine-grained sharing of personal data with privacy protection. 
\par
In order to realize personal data sharing autonomy, technically, 
a privacy-enhancing data-sharing system needs to, at least, 
achieve the abilities of \textit{selective sharing, informed consent-based 
authorization}, and \textit{permission revocation}. 
\textit{Selective sharing} 
is an approach to realize the ``minimization of data collection and processing'' and enables 
data owners to share their data 
(e.g., login credentials, transaction details, verification information) 
with selected service providers, 
providing only the information that is strictly needed by them, 
while specifying access times for the data. 
For example, in online forum applications, users can select any nickname 
    or account to post, and in instant messaging software, users can share data with 
    different friends and set the last accessible time of the data. 
It prevents the unwarranted disclosure of additional personal data, thereby 
protecting the privacy and control of the personal information. 
\textit{Informed consent-based authorization} requires that data consumers obtain 
explicit, informed consent from data owners prior to accessing, modifying, deleting, and forwarding their personal data. 
For example, when users first use the software, 
they must read the software's privacy policy in its entirety and click the ``agree'' button, 
and in online video conferences, participants must obtain consent 
from other attendees to record the meeting. 
This process ensures that data owners are fully aware 
of how their information will be processed. 
Since personal data is increasingly regarded as an individual's private property, 
explicit consent should be obtained before processing it. 
This paper focuses on the access aspect during personal data sharing. 
And \textit{permission revocation} 
is a compromised implementation of ``the right to be forgotten'' since the trusty erase 
of data is hard to realize\cite{tsesis2014right}. Permission revocation usually makes 
the data's decryption keys invalid in which data consumers cannot access the data anymore. 
\par
In fact, in order to protect personal data, numerous privacy-enhancing data sharing schemes have been 
proposed, which can be primarily categorized into two types: \textit{traditional encryption data sharing schemes} 
and \textit{attribute-based encryption (ABE) data sharing schemes}. \par
\textit{Traditional encryption data sharing schemes} use traditional 
cryptographic tools, e.g., symmetric encryption\cite{abdullah2017advanced} or  
public-key encryption\cite{canetti2003forward}, to encrypt data before storing it 
in cloud storage, thus ensuring 
confidentiality. However, such schemes fail to enable users to authorize 
multiple data consumers in a ``one-to-many'' access control manner. 
To address the above-mentioned issue, many traditional encryption data sharing schemes employ access control mechanisms such as role-based\cite{sandhu1998role} or 
identity-based\cite{boneh2001identity} access control to improve efficiency. 
Although these traditional data sharing systems achieve data confidentiality and 
``one-to-many'' data sharing, they fall short of the above abilities:  
selective sharing, informed-consent based authorization, and 
permission revocation. 
\par
\textit{Attribute-based encryption (ABE) data sharing schemes} have been proposed, 
firstly, to provide privacy-preserving, efficient, and fine-grained data sharing 
for data owners\cite{liang2009attribute},\cite{dong2014achieving},\cite{zuo2017fine},\cite{10098652},\cite{ning2020dual},
\cite{EMR},\cite{yang2022efficient}. 
The ABE\cite{FAME},\cite{riepel2022fabeo} 
enables personal data owners to provide their 
information to various data consumers 
whose attributes satisfy the access policy\cite{zhang2020attribute}. 
The scheme\cite{dong2014achieving} integrates ABE and identity-based encryption 
(IBE\cite{cocks2001identity}) to build a privacy-preserving data sharing system that 
supports the revocation of data consumers by updating part of the ciphertext. However, its 
revocation process lacks efficiency and granularity. 
And the scheme\cite{zuo2017fine} combines ABE and proxy re-encryption (PRE\cite{green2007identity},\cite{sun2022verifiable},\cite{zhang2023identity}) 
to design a two-factor protection mechanism to improve efficiency for 
decryption keys revocation of data consumers, but the scheme fails to allow the data owner to initiate the revocation process actively. 
Furthermore, the scheme\cite{EMR} proposes an electronic medical record sharing system 
integrating tamper resistance and supporting malicious user revocation without affecting honest users. 
Though the above-mentioned schemes realize the permission revocation in many ways, 
they fail to support selective sharing and informed-consent based authorization. 
\par
To realize selective sharing and informed-consent based authorization, 
the scheme\cite{yang2022efficient} combines searchable encryption (SE\cite{li2014enabling}) 
and ABE to design a selective data sharing system by 
selecting the keywords of data and sending 
them to data consumers. 
The keywords in the scheme\cite{yang2022efficient} represent the data owner's informed 
consent for data consumers to access their data. 
However, because of the separation between keywords and data, 
keywords can be forwarded by authorized consumers to unauthorized consumers maliciously 
without the data owners' authorization. 
For instance, in the scheme\cite{yang2022efficient}, attackers holding 
a keyword can collude with another attacker holding a set of attributes that satisfy 
the access policy to decrypt the ciphertext which is called collusion attacks\cite{zhang2020attribute}. \par
In addition, for efficient and convenient management of data, 
the above-mentioned schemes employ the cloud storage as the data keeper. 
However, they are unable to resist Economic Denial of Sustainability (EDoS\cite{sqalli2011edos},\cite{somani2015ddos}) attacks, 
where attackers excessively request the data of a data owner, 
leading to significant financial losses due to the overuse of cloud resources. 
To deal with this attack, 
the scheme\cite{ning2020dual} designs a dual access control for data sharing that 
enables sharing data securely and controls the download request 
of cloud resources, thus realizing EDoS resistance. However, it fails to realize 
personal data sharing autonomy. 
\par
In the above schemes, a single piece of data is not treated as the independent minimum 
unit for authorization and sharing, and the authorization to data consumers 
is not strictly and explicitly associated with the consent of data owners, mainly 
due to their design focus on the management of system administrators rather than 
enhancing the data owners' privacy. 
This leads to misuse of data by unauthorized data consumers, tampering with data by data keepers, and 
resulting in data owners losing control over their own data. 
To solve the above issues, we introduce a novel task-driven data capsule sharing system (TD-DCSS) realizing personal 
data sharing autonomy for flexible, secure, and privacy-enhancing data sharing. 
Data capsules\cite{wang2019data},\cite{gao2021teekap}, which store both 
the data and the access policy dictating their usage, offer the potential to act 
as the minimal units for independent and secure personal data storage and sharing. 
And the ``task-driven'' data sharing mechanism links a task to a consumer, 
with the task representing the owners' consent and the permission granted by the owner 
to the consumer.
The main contributions of the paper are stated as follows. 
\begin{itemize}
\item{\textit{Data capsule encapsulation method with tamper resistance. }
A method enabling personal data owners to encapsulate their data into 
data capsules is devised. 
Data capsules encapsulate the personal data, an access policy, and a 
bilinear pairing verification element. 
The proposed encapsulation method, building on the cumulative XOR random mask, 
bilinear pairing hash generation and ABE, generates a data capsule under an access policy. 
To protect the data capsules from tampering by semi-trusted data keepers, 
a bilinear pairing verification element, derived from the data capsule itself, 
is embedded within the capsule to ensure its integrity. 
}

\item{\textit{Task-driven data sharing mechanism with collusion and EDoS resistance. }
We propose a novel task-driven data sharing mechanism for 
selective data sharing, 
and formulate an efficient task construction methodology. 
The ``task-driven'' mechanism comprises two components: a task sent to data consumers and 
two tokens sent to the data keeper. The task, signifying the data owner's consent, 
includes the decryption keys for specific data granules. 
Since the task is associated with the data consumer's ID, the data consumer cannot 
collude with other data consumers to decrypt the data capsules. 
The two tokens, 
the download token and the revocation token, 
serve distinct purposes in our system. 
Specifically, preventing EDoS attacks is achieved by the download token, 
and revoking the data consumers' permissions is achieved by the revocation token. 
}

\item{\textit{Personal data sharing autonomy.}
The concept of personal data sharing autonomy is introduced, focusing on data 
owners' privacy. Personal data 
sharing autonomy, including selective sharing, informed-consent 
based authorization, and permission revocation, 
not only realizes the data privacy regulations: data collecting and processing minimization, explicit consent, 
and the right to be forgotten, 
but also enables data owners to have full control over their data. 
}
\item{\textit{Security and efficiency analysis.}}
We conduct a comprehensive security analysis, a theoretical analysis, 
and an experimental simulation. 
The security analysis shows that our scheme is correct, sound, and secure under 
the security model. The results of the theoretical analysis and 
the experimental simulation 
demonstrate that, compared 
with the state-of-the-art schemes, our scheme offers more advantageous features 
and stands as a highly efficient solution. 
\end{itemize}

\section{Overview and design goals}
In this section, we sketch the proposed data capsule encapsulation method and task-driven 
data sharing mechanism, providing intuition behind them. Then, we introduce the design goals 
that guide the development of the proposed scheme. In order to combine with the real scene, we model the role of service provider (SP) and 
cloud storage (CS) as data consumer and data keeper, respectively. \par
\subsection{Overview of the Proposed Scheme}
\begin{figure}[!t]
  \centering
  \includegraphics[width=3.4in]{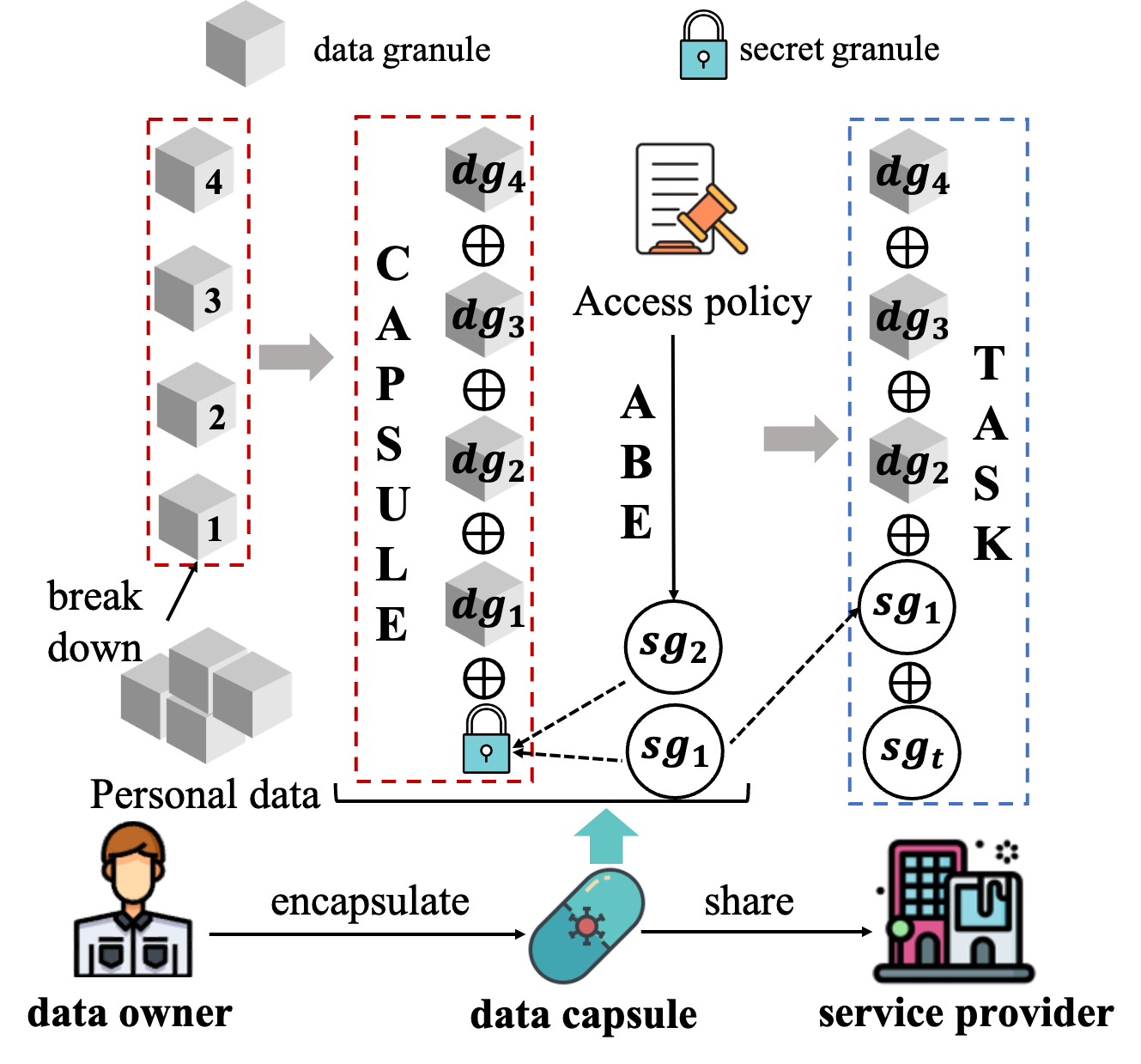}
  \caption{Data capsule encapsulation and sharing. }
  \label{fig:dc}
\end{figure}

Our proposed task-driven data capsule sharing system (TD-DCSS) scheme is based on a fast attribute-based mechanism 
(FABEO\cite{riepel2022fabeo}), and it can easily integrate with other CP-ABE schemes to support 
additional features (e.g., attribute revocation\cite{attrapadung2009attribute}, accountability\cite{10.1007/978-3-319-45741-3_28}, and policy hiding\cite{zhang2018security}). We introduce a data capsule encapsulation method and task-driven data 
sharing mechanism in TD-DCSS for flexible, efficient, and secure data sharing. The details are described as follows. 
\begin{itemize}
  \item \textit{Data capsule encapsulation method.} In real life, service providers may only need partial personal data from users to 
  successfully offer services. 
  For instance, when a hospital verifies a user's age, 
  it is sufficient to provide the information from the ID related to the birth date.
  Providing the entire ID in this situation may pose a privacy risk.
  To ease this risk, we propose a data capsule encapsulation method as shown in Fig. \ref{fig:dc}. 
  The main idea is that data owners can initially break down their data into granules, 
  which can be shared in the future upon demand. Subsequently, owners can XOR all the 
  data granules, including a secret granule generated under an access policy through encryption operations in ABE. 
  Finally, the data capsule (DC) comprises a set of data granules 
  $dg_1, dg_2, dg_3, dg_4$ and a secret granule $sg$, which is generated under the 
  specified access policy. Specifically, the data capsule is represented as 
  $dg_1 \oplus dg_2 \oplus dg_3 \oplus dg_4 \oplus sg$. 
  \item \textit{Task-driven data sharing mechanism}. 
  To realize selective sharing and informed-consent based authorization, we 
  propose a novel task-driven data sharing mechanism, and the main idea is 
  shown in Fig. \ref{fig:dc}. 
  When the data owner wishes to share $dg_1$ in the data capsule selectively
  with the service provider, 
  it issues a task and sends the task to the SP as the consent for accessing its data. 
  Specifically, the $sg$ consists of two distinct components, $sg_1$ and $sg_2$, where $sg_1$ is randomly 
  generated, while $sg_2$ is generated under the specified access policy. 
  In this scenario, the task takes the form of $dg_2 \oplus dg_3 \oplus dg_4 \oplus sg_1 \oplus sg_t$, 
  where $sg_t$ is associated with the target ID of the SP. 
  After retrieving the data capsule from the CS and receiving the task from the owner, 
  the SP can recover $dg_1$ by reconstructing the $sg_2$ and $sg_t$, and then computing 
  $dg_1=\text{DC}\oplus\text{task}\oplus sg_2 \oplus sg_t$. 
\end{itemize}
\subsection{Design Goals}
  Our goal is to build a task-driven data capsule sharing system, 
  realizing personal data sharing autonomy, for flexible, secure, and privacy-enhancing data sharing. There are four key 
  objectives in our scheme: correctness, soundness, security, and efficiency. 
  \begin{itemize}
    \item \textit{Correctness}: A service provider with a data capsule encrypted 
    under an access policy, a task bound to its ID, and its attributes that satisfy 
    the access policy, can correctly recover the data encrypted in the data capsule 
    and task.
    \item \textit{Soundness}: Service providers should refrain from downloading 
    data capsules from cloud storage when the task issuer is wrong, 
    and should not receive tasks or data capsules that cannot be decrypted. The cloud 
    storage should not learn any sensitive information about the data capsule when updating 
    it. 
    \item \textit{Security}: Unauthorized service providers and cloud storage 
    should have no knowledge of personal data. The task must not disclose 
    any information beyond what is contained in both the data capsule 
    and the task itself.
    \item \textit{efficiency}: The computation and storage consumption of the scheme 
    should not be too much. 
  \end{itemize}
    
\section{Preliminaries}
\subsection{Access Structure}
  Our access structure is denoted by $\mathbb{A}=(\mathbb{M},\pi)$, where policy matrix $\mathbb{M}\subseteq \mathbb{Z}_p^{n_1\times n_2}$ 
  and mapping $\pi:[n_1]\rightarrow \mathcal{U}$, where $[n]$ denotes the set $\{1,2,...,n\}$, and $\mathcal{U}$ denotes the universe 
  of attributes. 
  Let $\mathcal{S}\models \mathbb{A}$ 
  indicate that $\mathcal{S}$ satisfies the access policy 
  ($\mathbb{A}=\mathbb{M}, \pi$), where $\mathcal{S}\in \mathcal{U}$ denotes a set of attributes. 
  We use the same definitions $\rho(i)= |\{z|\pi(z)=\pi(i),z\leq i\}|$ 
  and $\tau=\ $max$_{i\in [n_1]}\rho(i)$ corresponding to the maximum number of times 
  an attribute is used in $\mathbb{M}$ as \cite{riepel2022fabeo}. 
  Note that any boolean formula can be transferred to $(\mathbb{M},\pi)$ in polynomial time\cite{lewko2011unbounded}. 
  \subsection{Bilinear Pairing}\label{section:bp}
  Let $\lambda$ be a security parameter and GroupGen be a probabilistic polynomial time algorithm that takes the 
  $1^{\lambda}$ and outputs group description $(p,\mathbb{G}_1,\mathbb{G}_2,\mathbb{G}_T,e:
  \mathbb{G}_1\times \mathbb{G}_2\rightarrow \mathbb{G}_T,g_1,g_2)$, where 
  p is the group order of $\mathbb{G}_1,\mathbb{G}_2$ and $\mathbb{G}_T$, and $g_1,g_2$ is 
  the generator of $\mathbb{G}_1,\mathbb{G}_2$ respectively. The mapping e has the following 
  properties:
  \begin{itemize}
    \item \textbf{Bilinearity:} For all $u\in\mathbb{G}_1,v\in\mathbb{G}_2$ and $x,y \in\mathbb{Z}_p$, $e(u^x,v^y)=e(u,v)^{xy}$
    \item \textbf{Non-degeneracy:} $\exists u\in\mathbb{G}_1,v\in\mathbb{G}_2$, such that $e(u^x,v^y)=I_{\mathbb{G}_T}$, where $I_{\mathbb{G}_T}$ 
    denotes the identity element of $\mathbb{G}_T$. 
    \item \textbf{Computability:} $e$ can be easily and efficiently computed. 
  \end{itemize}
  \subsection{Security Assumption}
  \textit{Definition 1 (DBDH):} Let $\mathcal{BG}$ be a type-\uppercase\expandafter{\romannumeral3} 
  pair ($\mathbb{G}_1,\mathbb{G}_2,\mathbb{G}_T,p,e$) with generators $g\in\mathbb{G}_1,h\in\mathbb{G}_2$. 
  Given a DBDH tuple ($g,g^a,g^b,g^c,h,h^a,h^b,h^c,\mathcal{Z}$), where 
  $a,b,c \in \mathbb{Z}_p$, the goal of the decisional bilinear Diffie-Hellman (DBDH) problem is to 
  distinguish whether $\mathcal{Z}=e(g,h)^{abc}$ or $\mathcal{Z}=\mathcal{R}$, where 
  $\mathcal{R}$ is a random element of $\mathbb{G}_T$. 
  \subsection{Notation}
    For the convenience, the frequently used symbols are 
    presented in TABLE \ref{tab:notations}.
  \begin{table}[!t]
    \caption{Frequently Used Notations 
    \label{tab:notations}}
    \centering
    \begin{tabular}{cl}
    \hline
    \textbf{\textit{Notation}} & \textbf{\textit{Description}}\\
    \hline
    $\lambda$ & \textit{System security parameter}\\
    $\ell$ & \textit{The length of data granules}\\
    $\mathcal{DG}_{n}$ & \textit{A set of $n$ data granules}\\
    $\mathbb{A}=(\mathbb{M},\pi)$ & \textit{An access policy}\\
    $\mathcal{T},\mathcal{R},\mathcal{D}$ & \textit{a task, revocation token and download token}\\
    $\mathcal{L}$ & \textit{The local secret data for issuing tasks}\\
    $\mathcal{I}$ & \textit{A set of index of data granules in $\mathcal{DG}_{n}$}\\
    $h,\{\mathcal{H}_i\}_{i\in[3]}$ & \textit{The efficient hash function used in system}\\
    \hline
    \end{tabular}
  \end{table}
\section{Definition}\label{section:definition}
  \subsection{Task-Driven Data Capsule Sharing System}\label{set:3}
    \textit{Definition 2 (TD-DCSS): The task-driven data capsule sharing system consists
    of the following algorithms:}
    \begin{itemize}
        \item Setup($1^{\lambda})\rightarrow(mpk,msk)$\textit{: The Setup algorithm takes a system security 
        parameter $\lambda\in \mathbb{N}$ as input, and outputs a master public key $mpk$ and a master secret key $msk$.}
        \item KeyGenSP($mpk,msk,ID_{SP},\mathcal{S}\subseteq \mathcal{U})\rightarrow sk_{SP}$\textit{: 
            The KeyGenSP algorithm takes the master public key $mpk$, the master secret key $msk$, an 
            $ID_{SP}$ of the target service provider and an attribute set $\mathcal{S}\subseteq \mathcal{U}$ 
            as input, and outputs secret key $sk_{SP}$ of the SP. 
        }
        \item GenSeed($mpk,ID_{PDO})\rightarrow (\gamma,\psi)$\textit{: 
            The GenSeed algorithm takes the master public key $mpk$ and an $ID_{PDO}$ of the personal data owner as input, 
            and outputs a pair of seed ($\gamma,\psi$). 
        }
        \item PKeyGenPDO($\psi)\rightarrow (pk_{PDO},\beta)$\textit{: 
            The PKeyGenPDO algorithm takes the seed $\psi$ as input, 
            and outputs a factor $\beta$ as well as a $pk_{PDO}$.
        }
        \item SKeyGenPDO($\gamma,\beta)\rightarrow sk_{PDO}$\textit{: 
            The SKeyGenPDO algorithm takes the seed $\gamma$ and the factor $\beta$ as input, 
            and outputs a secret key $sk_{PDO}$ of the PDO.
        }
        \item Encapsulate($mpk,sk_{PDO},\mathcal{DG}_{n},\mathbb{A})\rightarrow (DCI,\mathcal{L},DC)$\textit{: 
            The Encapsulate algorithm takes the master public key $mpk$, the secret key $sk_{PDO}$, 
            a set of $n$ data granules $\mathcal{DG}_{n}$ and an access policy $\mathbb{A}$ as input, 
            and outputs the data capsule identifier $DCI$, a local secret parameter $\mathcal{L}$ 
            and a data capsule $DC$ corresponding to the $DCI$. 
        }
        \item TaskIssue($mpk,sk_{PDO},ID_{SP},\mathcal{DG}_{n},\mathcal{I},\mathcal{L},t)\rightarrow (\mathcal{T},\mathcal{R},$ $\mathcal{D})$\textit{: 
            The Taskissue algorithm takes the master public key $mpk$, the secret key $sk_{PDO}$, 
            an $ID_{SP}$, a set of $n$ data granules $\mathcal{DG}_{n}$, a set of indices $\mathcal{I}$ 
            for $\mathcal{DG}_{n}$\footnote{$\mathcal{I}$ represents a set of indices of data granules in $\mathcal{DG}_{n}$ to be shared with the SP. }, 
            a local secret parameter $\mathcal{L}$ corresponding to $\mathcal{DG}_{n}$ 
            and a task expiry time $t$ as input, and outputs a task $\mathcal{T}$, a revocation token $\mathcal{R}$ and 
            a download token $\mathcal{D}$. 
        }
        \item AccessDC($mpk,sk_{SP},DCI,\mathcal{T},pk_{PDO})\rightarrow P_{\mathcal{T},1}$\textit{: 
            The AccessDC algorithm takes the master public key $mpk$, the secret key $sk_{SP}$, 
            a data capsule identifier $DCI$, a task $\mathcal{T}$ and the $pk_{PDO}$ that issues 
            the $\mathcal{T}$ as input, and outputs a download parameter $P_{\mathcal{T},1}$. 
        }
        \item DownloadDC($DCI,\mathcal{D},P_{\mathcal{T},1})\rightarrow DC$\textit{: 
            The DownloadDC algorithm takes the data capsule identifier $DCI$, a download token $\mathcal{D}$ 
            corresponding to $DCI$ and a download parameter $P_{\mathcal{T},1}$ as input, and outputs the $DC$.
        }
        \item DecDC($mpk,sk_{SP},DCI,DC,\mathcal{T},P_{\mathcal{T},1})\rightarrow \{dg_w\}_{w\in\mathcal{I}}$\textit{: 
            The DecDC algorithm takes the master public key $mpk$, the secret key $sk_{SP}$, 
            a data capsule identifier $DCI$, a data capsule $DC$, a task $\mathcal{T}$ and 
            a download parameter $P_{\mathcal{T},1}$ as input, and outputs a set of data granules $\{dg_w\}_{w\in\mathcal{I}}$. 
        }
        \item UpdateDC($mpk,DCI,DC,\mathcal{R})\rightarrow (DCI',DC')$\textit{: 
            The UpdateDC algorithm takes the master public key $mpk$, a data capsule identifier $DCI$, a data capsule $DC$ 
            and the revocation token $\mathcal{R}$ as input, and outputs an updated $DCI'$ and an updated $DC'$. 
        }
    \end{itemize}
  \subsection{Definition of Correctness}
  The correctness of TD-DCSS is that when a service provider possesses a task corresponding to the data capsule and its attributes satisfy the access policy linked to the data capsule, it can successfully recover the data encrypted in both the data capsule and the task. \par
  \textit{Definition 3 (Correctness):} Our TD-DCSS scheme is correct, if $\forall\lambda\in\mathbb{N}$ and $\mathcal{S}\models \mathbb{A}$, we have 
  \[\small \textbf{Pr}[\text{DecDC}(mpk,sk_{SP},DCI,DC,\mathcal{T},P_{\mathcal{T},1})=\{dg_w\}_{w\in\mathcal{I}}]=1\]
  where $P_{\mathcal{T},1}=\text{AccessDC}(mpk,sk_{SP},DCI,\mathcal{T},pk_{PDO})$ and 
  the probability is taken with respect to the choice of 
  \begin{align*}
    (mpk,msk) &\leftarrow \text{Setup}(1^{\lambda}),\\
    (\gamma,\psi)&\leftarrow \text{GenSeed}(mpk,ID_{PDO}),\\
    sk_{PDO},pk_{PDO} &\leftarrow \text{PKeyGenPDO}(\psi),\text{SKeyGenPDO}(\gamma,\beta),\\
    sk_{SP} &\leftarrow \text{KeyGenSP}(mpk,msk,ID_{SP},\mathcal{S}\subseteq \mathcal{U}),\\
    (DCI,\mathcal{L},DC) &\leftarrow \text{Encapsulate}(mpk,sk_{PDO},\mathcal{DG}_{n},\mathbb{A}),\\
    (\mathcal{T},\mathcal{R},\mathcal{D})&\leftarrow \text{TaskIssue}(mpk,sk_{PDO},ID_{SP},\mathcal{DG}_{n},\mathcal{I},\mathcal{L},t).\\
  \end{align*}\par
  The correctness of the scheme ensures that authorized parties can recover the data as intended. 
  \subsection{Definition of Soundness}
  In TD-DCSS,  
  we define the soundness to cover the following three cases 
  in which system participants do not act as expected, 
  which are not addressed in security. 
  \begin{itemize}
    \item \textit{Case 1 (Soundness of AccessDC)}: If the secret key inside the task does 
    not match the public key inside the AccessDC algorithm, 
    or the factor in $\mathcal{L}$ inside the task does not match the factor in $\mathcal{L}$ 
    inside the data capsule identifier $DCI$, the AccessDC algorithm would be unable to 
    correctly recover the download parameter. 
    \item \textit{Case 2 (Soundness of DecDC)}: The download parameter and the result of 
    ABE-related decryption are tied to the ID of an SP. This means that the 
    probability of an SP holding a task colluding with another SP, 
    whose attributes satisfy the access policy, to recover the data granules 
    successfully is negligible.  
    \item \textit{Case 3 (No Leakiness of UpdateDC)}: The revocation token is sent to 
    the CS, which possesses partial information about the local parameter 
    $\mathcal{L}$. However, the CS lacks complete knowledge of the parameters 
    within $\mathcal{L}$, which means that the CS can successfully recover 
    the important parameters inside $\mathcal{L}$ with a negligible probability. 
  \end{itemize}
  \subsection{Definition of Security}
  The formal definition of security for the TD-DCSS scheme is described as follows. \par
  \textit{Definition 4 }(${\mathsf{IND}\text{-}\mathsf{CPA}}$): The ${\mathsf{IND}\text{-}\mathsf{CPA}}$ security model of 
  $\mathcal{DCSS}_{\mathcal{TD}}$ can be described through a game involving an adversary $\mathcal{A}$ and 
  a challenger $\mathcal{C}$. \par
  \textbf{Init. } $\mathcal{A}$ submits a challenge access 
  policy $\mathbb{A}^*$ to $\mathcal{C}$.\par 
  \textbf{Setup.} With a system security parameter $\ell \in \mathbb{N}$, $\mathcal{C}$ 
  runs the system setup algorithm Setup($1^{\lambda}$) to generate a master public key 
  $mpk$, a master secret key $msk$, and initializes the universe of attributes 
  $\mathcal{U}$. $\mathcal{C}$ sends $mpk$ and $\mathcal{U}$ to $\mathcal{A}$ 
  while securely retaining the $msk$. \par
  \textbf{Phase 1.} $\mathcal{A}$ can adaptively send a sequence of the following 
  queries to $\mathcal{C}$. 
  \begin{itemize}
    \item $\mathcal{O}_{\text{KSP}}(ID_{SP},\mathcal{S}\subseteq \mathcal{U})$: 
    $\mathcal{A}$ can submit a query to the key generation of SP oracle with the 
    input message containing an $ID_{SP}\in \{0,1\}^*$, 
    and a set of attributes $\mathcal{S} \subseteq \mathcal{U}$. 
    If $\mathcal{S}\models \mathbb{A}^*$, $\mathcal{C}$ outputs a $\bot$ instead. Otherwise, 
    $\mathcal{C}$ runs the key generation of SP algorithm, 
    KeyGenSP($mpk, msk, ID_{SP}, \mathcal{S} \subseteq \mathcal{U}$), 
    to generate the secret key $sk_{SP}$ and sends it to $\mathcal{A}$. 
    \item $\mathcal{O}_{\text{GenSeed}}(ID_{PDO})$: 
    $\mathcal{A}$ can submit a query to the seed generation oracle 
    with the input message containing an identifier $ID_{PDO}\in \{0,1\}^*$. 
    $\mathcal{C}$ runs the seed generation algorithm, 
    GenSeed($mpk,ID_{PDO}$), 
    to generate two seeds $\gamma,\psi$ and sends $\psi$ to $\mathcal{A}$. 
    \item $\mathcal{O}_{\text{PKPDO}}(\psi)$: 
    $\mathcal{A}$ can submit a query to the public key generation of PDO oracle 
    with the input message containing a group element $\psi\in\mathbb{G}_2$. 
    $\mathcal{C}$ runs the public key generation of PDO algorithm, 
    PKeyGenPDO($\psi$), 
    to generate a public key $pk_{PDO}$, a factor $\beta$ and 
    sends $pk_{PDO},\beta$ to $\mathcal{A}$. 
    \item $\mathcal{O}_{\text{SKPDO}}(ID_{PDO},\beta)$: 
    $\mathcal{A}$ can submit a query to the public key generation of PDO oracle 
    with the input message containing an $ID_{PDO} \in \{0,1\}^*$ and a 
    factor $\beta \in \mathbb{Z}^*_p$. 
    $\mathcal{C}$ runs the secret key generation of PDO algorithm, 
    SKeyGenPDO($\gamma,\beta$), 
    to generate a secret key $sk_{PDO}$ and 
    sends $sk_{PDO}$ to $\mathcal{A}$. 
  \end{itemize}\par
  In phase 1, 
  when $\mathcal{A}$ queries to generate a pair of public key and secret key of PDO, 
  the seed generation oracle $\mathcal{O}_{\text{GenSeed}}(ID_{PDO})$, 
  the public key generation oracle $\mathcal{O}_{\text{PKPDO}}(\psi)$, 
  and the secret generation oracle $\mathcal{O}_{\text{SKPDO}}(\gamma,\beta)$ 
  must be queried in a sequential order. \par
  \textbf{Challenge.} $\mathcal{A}$ submits two sets of data granules ($m_0, m_1$) 
  with the same length to $\mathcal{C}$. 
  $\mathcal{C}$ flips a random coin 
  $b\in\{0,1\}$. $\mathcal{C}$ then 
  runs the data capsule encapsulation algorithm 
  Encapsulate($mpk,\cdot,m_{b},\mathbb{A}^*)$, to generate 
  ($DCI^*,DC^*,\mathcal{L}^*$) and sends $DCI^*,DC^*$ 
  to $\mathcal{A}$. \par
  \textbf{Phase 2.} Under the previous restrictions, 
  $\mathcal{A}$ continues to query $\mathcal{C}$. 
   \par
  \textbf{Guess.} $\mathcal{A}$ outputs a guess $b'$. If $b'=b$, $\mathcal{A}$ wins 
  the game. The advantage of $\mathcal{A}$ in this game is defined as 
  \[\mathbf{Adv}^{{\mathsf{IND}\text{-}\mathsf{CPA}}}_{\mathcal{DCSS}_{\mathcal{TD}},\mathcal{A}}(\lambda)=|\mathbf{Pr}[b=b']-\frac{1}{2}|\]
  \section{System Architecture}
  \begin{figure}[!t]
    \centering
    \includegraphics[width=3in]{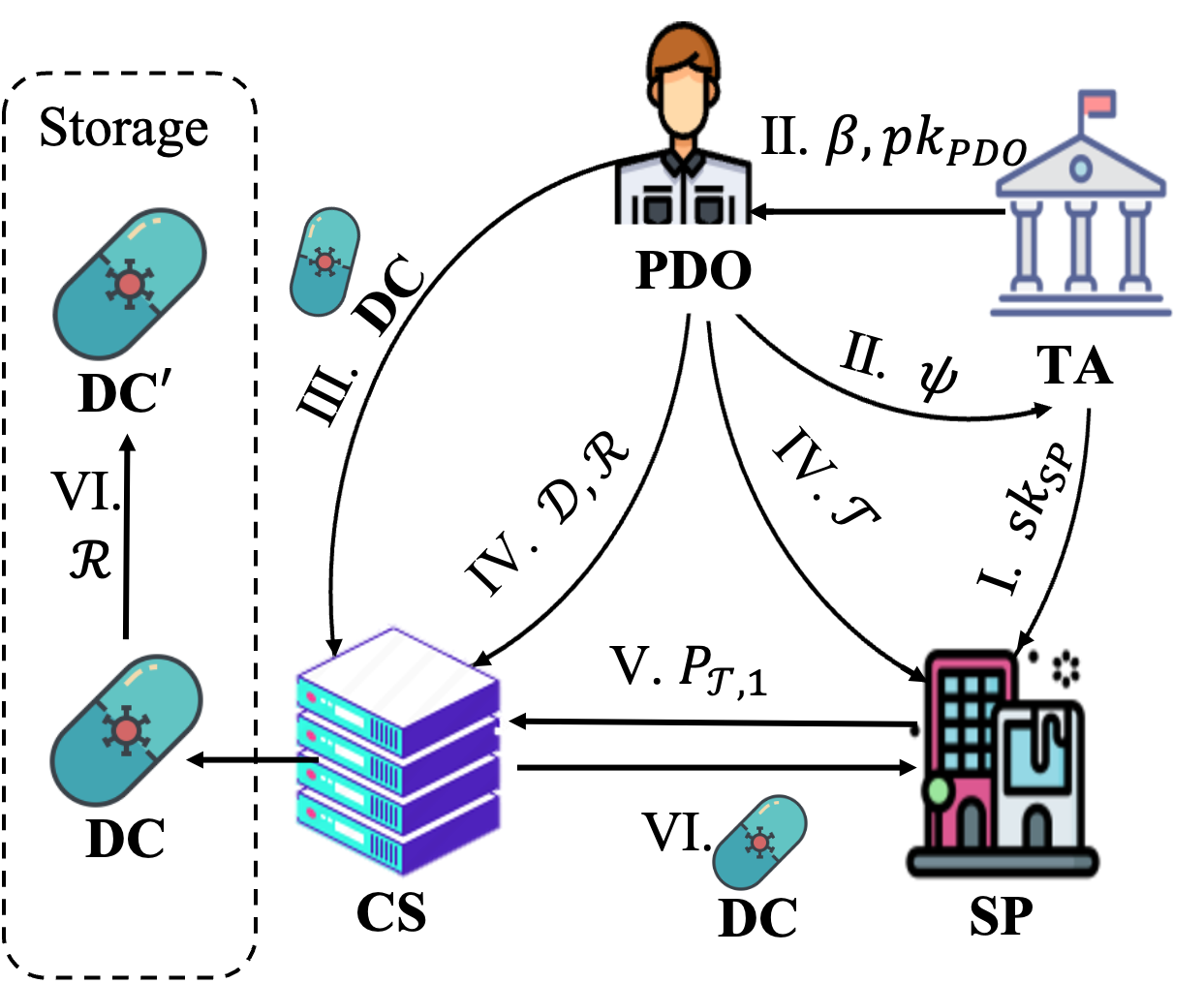}
    \caption{System model. }
    \label{fig:system-model}
  \end{figure}
\subsection{System Model}
  The Fig. \ref{fig:system-model} describes a task-driven data capsule sharing system (TD-DCSS), 
  in which personal data owners (PDOs) want to share 
  personal data with multiple service providers 
  for utilization of their services. 
  PDOs encapsulate their data into data capsules (DCs) under an access policy and upload DCs to the cloud storage (CS). 
  PDOs selectively share their data and grant authorization to service providers (SPs)
  by issuing tasks and sending tasks to SPs. 
  Since PDOs require that the task be a one-time token, 
  the CS revokes the permissions of SPs after they use the task by updating part of 
  the data capsules. 
  A TA, a CS, PDOs, and SPs as system entities are given the 
  details below. 
  \begin{enumerate}
    \item{
      \textit{TA}: The trusted authority (TA), 
      such as the government, bears responsibility for initiating the system 
      parameters, the universe of attributes 
      and issues the secret keys to SPs 
      (see \uppercase\expandafter{\romannumeral1} in Fig. \ref{fig:system-model}) 
      as well as the public keys to PDOs (see \uppercase\expandafter{\romannumeral2} in Fig. \ref{fig:system-model}).
    }
    \item{
      \textit{CS}: The cloud storage (CS) as a semi-trusted entity 
      provides abundant storage capacity and is responsible for archiving data capsules 
      (see \uppercase\expandafter{\romannumeral3} in Fig. \ref{fig:system-model}) and the associated 
      revocation/download tokens (see \uppercase\expandafter{\romannumeral4} in Fig. \ref{fig:system-model}). It  
      updates a data capsule according to the tokens provided by 
      a PDO (see \uppercase\expandafter{\romannumeral6} in Fig. \ref{fig:system-model}) and rejects the request from expired tasks. 
    }
    \item{
      \textit{PDO}: Personal data owners (PDOs) as the predominant participants 
      in the system can break down their data into data granules, 
      subsequently encapsulate these granules into a data capsule, 
      and outsource it to the CS (see \uppercase\expandafter{\romannumeral3} in Fig. \ref{fig:system-model}). 
      Furthermore, PDOs can issue tasks to the designated service providers (see \uppercase\expandafter{\romannumeral4} in Fig. \ref{fig:system-model}) 
      and submit the download token as well as the revocation token to the CS (see \uppercase\expandafter{\romannumeral4} in Fig. \ref{fig:system-model}). 
    }
    \item{
      \textit{SP}: The service providers (SPs) who have many attributes and 
      provide service for PDOs are the important system participants. 
      They accept tasks (see \uppercase\expandafter{\romannumeral5} in Fig. \ref{fig:system-model}) and decrypt the data capsules to access personal 
      data that is necessitated for providing service. 
    }
  \end{enumerate}
  \subsection{System Phases}
  \begin{figure}[!t]
    \centering
    \includegraphics[width=3in]{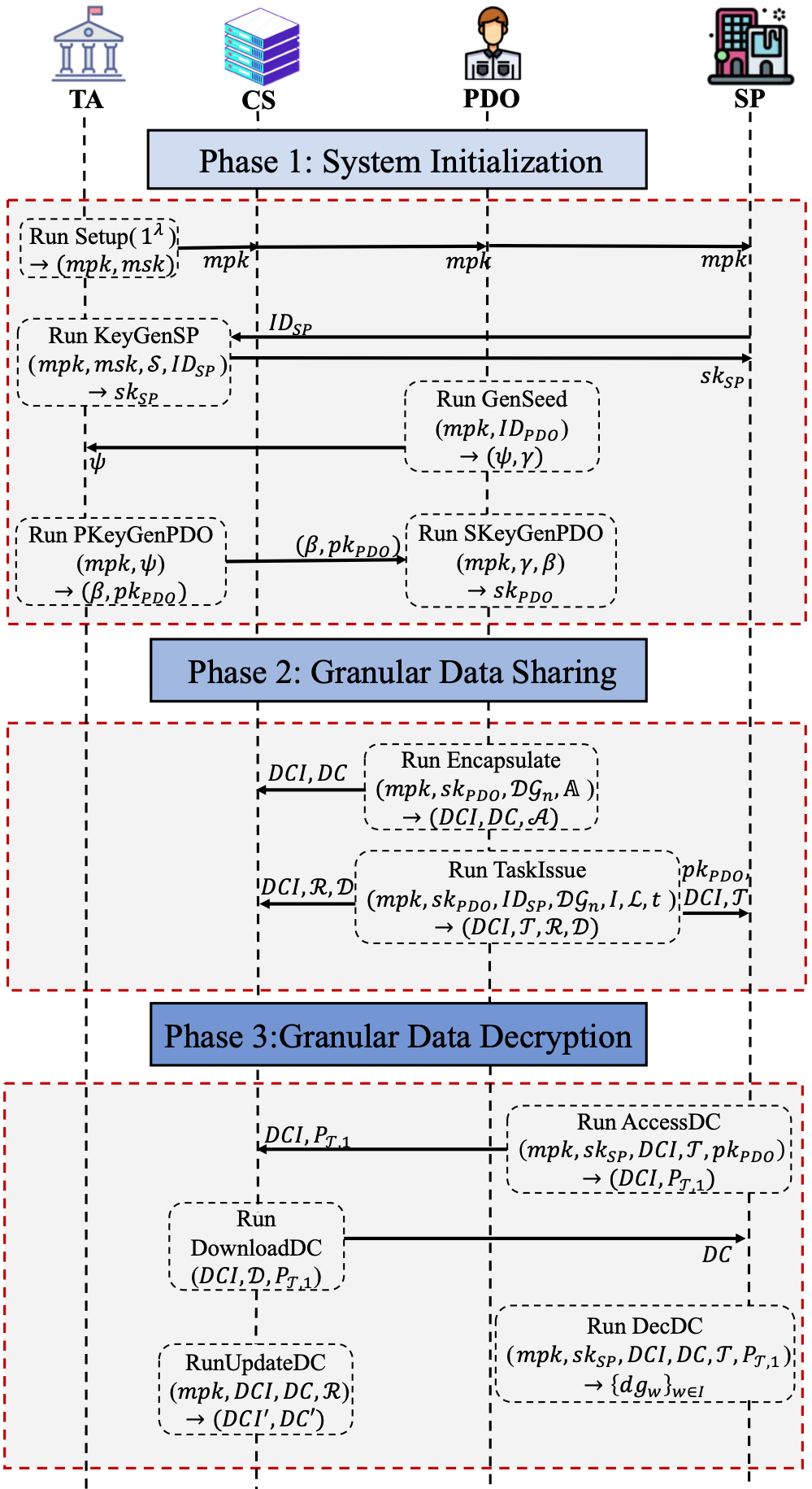}
    \caption{Data sharing phases in our system.}
    \label{fig:system-flows}
  \end{figure}
 Our TD-DCSS scheme has the following 
    several functions: Setup, KeyGenSP, GenSeed, PKeyGenPDO, SKeyGenPDO, Encapsulate, TaskIssue, AccessDC, DownloadDC, 
    DecDC, and UpdateDC, which are defined in Section \ref{set:3}. 
    These functions are used in the following phases: System Initialization, Granular Data Sharing, 
    and Granular Data Decryption. As shown in Fig.\ref{fig:system-flows}, 
    the three phases are described as follows: 
\begin{enumerate}
    \item {
        \textit{System Initialization}: As shown in Fig. \ref{fig:system-flows}, 
        the TA generates the master public key $mpk$ and master secret key $msk$ 
        by running Setup($1^{\lambda}$). 
        TA runs KeyGenSP($mpk,msk,ID_{SP},\mathcal{S}\subseteq \mathcal{U})$ 
        with an $ID_{SP}$ and a set of attributes $\mathcal{S}\subseteq \mathcal{U}$ to 
        generate a secret key $sk_{SP}$ for each SP in the system, where $\mathcal{U}$ 
        is the universe of attributes. 
        PDO runs GenSeed($mpk,ID_{PDO})$ with its 
        $ID_{PDO}$ to get a pair of seed $\gamma$, $\psi$. 
        Then, PDO sends $\psi$ to TA. After 
        receiving the $\psi$, TA selects a random mask $\beta\in\mathbb{Z}_p$ and runs 
        PKeyGenPDO($\psi)$ to get $pk_{PDO}$ and sends $\beta,pk_{PDO}$ to PDO. 
        Then, PDO runs SKeyGenPDO($\gamma,\beta)$ to generate $sk_{PDO}$. 
    }
    \item {
        \textit{Granular Data Sharing}: As shown in Fig. \ref{fig:system-flows}, to encapsulate the personal data into a data capsule,
        PDO first selects an access policy $\mathbb{A}$. Next, PDO 
        breaks down personal data into data granules $\mathcal{DG}_n$ with length $\ell$. 
        Then, PDO runs Encapsulate($mpk,sk_{PDO},\mathcal{DG}_{n},\mathbb{A})$ 
        to generate the data capsule identifier $DCI$, the data capsule $DC$, and the local secret parameter $\mathcal{L}$ which is stored locally. 
        Finally, PDO sends $DCI,DC$ to the CS. The CS stores 
        $DC$ indexed by $DCI$. 
        When sharing with service providers (SPs), PDO 
        runs TaskIssue($mpk,sk_{PDO},ID_{SP},\mathcal{DG}_{n},\mathcal{I},\mathcal{L},t)$ to 
        obtain a task $\mathcal{T}$, a revocation token $\mathcal{R}$, and a download token $\mathcal{D}$, 
        with the input $sk_{SP}$, the data granules $\mathcal{DG}_n$, the $ID_{SP}$ of the target SP, an expiry time for task $t$, 
        and a set of indices of data granules in $\mathcal{DG}_n$ to be shared with the SP. 
        Then PDO sends $pk_{PDO},DCI,\mathcal{T}$ to the SP, 
        and sends $DCI,\mathcal{R},\mathcal{D}$ to the CS. 
    }
    \item {
        \textit{Granular Data Decryption}: As shown in Fig. \ref{fig:system-flows}, after  
        receiving $pk_{PDO},DCI,\mathcal{T}$, 
        SP first runs AccessDC($mpk,sk_{SP},DCI,\mathcal{T},pk_{PDO})$ to obtain the 
        download parameter $P_{\mathcal{T},1}$ and sends it to the CS. 
        The CS checks the validity and timeliness of $P_{\mathcal{T},1}$ by running 
        DownloadDC($DCI,\mathcal{D},P_{\mathcal{T},1})$. 
        If DownloadDC returns $DC$, the CS sends it to the SP. Meanwhile, 
        the CS runs UpdateDC($mpk,DCI,\mathcal{R})$ with $\mathcal{R}$ provided by the 
        PDO to immediately update the $DC$ and revoke the permission of the SP. 
        With the received $DC$, the SP recovers the data granules by running 
        DecDC($mpk,sk_{SP},DCI,DC,\mathcal{T},P_{\mathcal{T},1})$. 
    }
\end{enumerate}
\subsection{Threat Model}
In our task-driven data capsule sharing system, the TA and 
PDOs have full trust. The TA honestly sets up the system parameters, 
initializes the universe of attributes, 
issues secret keys to SPs, and publishes the public keys of PDOs. 
PDOs can encapsulate their data into data capsules, 
outsource data capsules into the CS, and issue tasks to SPs for selectively 
sharing their data with them through our proposed scheme.	
\par
The CS, as a semi-trusted party, has huge storage resources to store data capsules 
from PDOs. The CS also assists PDOs in updating the data capsules using the revocation 
token from PDOs and runs access control over data capsules when SPs access data 
capsules using a download parameter. However, the CS may try to learn sensitive 
information from data capsules, the revocation token, and the download 
token individually. \par
The SPs are untrusted and may try to learn unauthorized information from data capsules and tasks. 
Furthermore, the SPs can be categorized into three types: type 1, type 2, and type 3, 
where type 1 of SPs have the correct task but their attributes do not satisfy the access 
policy of the data capsules, type 2 of SPs have attributes that satisfy the access policy 
of data capsules but do not have the correct tasks, and type 3 of SPs do not have a 
correct task and their attributes also do not satisfy the access policy of the data 
capsules. \par
Moreover, a collusion attack in our system occurs when type 1 and type 2 of SPs 
combine their attributes and tasks to access data granules within data capsules 
that they cannot obtain independently.
\par
The soundness and security model of TD-DCSS, as shown in the definition of soundness and 
security (Section \ref{section:definition}), formalize the above threat model, respectively. 
Specifically, in the $\mathsf{IND}\text{-}\mathsf{CPA}$ 
model and the soundness, 
the adversary simulates the CS and the SPs as follows:	
\begin{itemize}
  \item \textit{Untrusted service providers.} 
  The adversary can get many secret keys 
  by querying the secret key generation of the SP oracle many times, which enables the 
  adversary to simulate the three types of SPs. 
  Specifically, the adversary can act as type 1 of SPs that hold a correct task but the 
  attribute set of secret keys does not satisfy the access policy. 
  And the adversary can act as type 2 of SPs, if the attribute set associated with the issued secret keys 
  satisfies the challenge message but does not have a task. 
  Moreover, the adversary can act as 
  type 3 of SPs, if the attribute set associated with the issued secret keys do not 
  satisfy the challenge message and do not have a correct task. 
  After the adversary gets secret keys, 
  the adversary can obtain a ciphertext in the challenge phase and update the ciphertext in phase 2. 
  The adversary wins the $\mathsf{IND}\text{-}\mathsf{CPA}$ game by correctly guessing the bits of the underlying message.	
  \item \textit{Semi-trusted CS.} The adversary can obtain the original ciphertext and 
  the updated ciphertext without querying the secret key generation of the SP oracle. 
  The adversary wins the $\mathsf{IND}\text{-}\mathsf{CPA}$ game by guessing the underlying message of the 
  ciphertext correctly. 
\end{itemize}
\begin{figure*}[!htbp]
  \begin{tcolorbox}[colback=white,colframe=black,arc=0pt,outer arc=0pt,width=\textwidth]
    \raggedright  
    \underline{Setup($1^\lambda$)}: Initialize the universe of attributes $\mathcal{U}$. Run GroupGen($1^\lambda$) 
    to obtain $ (p,\mathbb{G}_1,\mathbb{G}_2,\mathbb{G}_T,e,g_1,g_2)$, where $e:\mathbb{G}_1\times \mathbb{G}_2 \rightarrow \mathbb{G}_T$ is a non-degenerate bilinear map (Section \ref{section:bp}). Pick $\alpha \in \mathbb{Z}_p$. Select the following hash functions, 
      \[
      h:\{0,1\}^*\rightarrow \mathbb{Z}_p^*, \quad
      \mathcal{H}_1:\{0,1\}^* \rightarrow \mathbb{G}_1, \quad 
      \mathcal{H}_2:\mathbb{G}_T \rightarrow \{0,1\}^{\ell}, \quad \mathcal{H}_3:\mathbb{G}_2^2\times \{0,1\}^{\ell}\times \mathbb{G}_1^{\tau +n_1} \rightarrow \mathbb{Z}_p^*,
      \]
      where $n_1$ is the number of rows in the policy matrix $\mathbb{M}$ and $\tau=\ $max$_{i\in [n_1]}\rho(i)$, where $\rho(i)= |\{z|\pi(z)=\pi(i),z\leq i\}|$ and $\pi:[n_1]\rightarrow \mathcal{U}$. Output the master public key and master private key as
      \[mpk=(p,\mathbb{G}_1,\mathbb{G}_2,\mathbb{G}_T,e,g_1,g_2,g_2^{\alpha},h,\{\mathcal{H}_i\}_{i\in [3]}),
        \qquad
        msk = \alpha.
      \]
      \underline{KeyGenSP($mpk,msk,ID_{SP},\mathcal{S}\subseteq \mathcal{U}$)}: 
      Pick $r\in \mathbb{Z}_p$. For each $s\in \mathcal{S}$, compute $sk_{1,s}=\mathcal{H}_1(s)^r$. 
      Parsing $msk=\alpha$, compute 
      \[sk_2=\mathcal{H}_1(ID_{SP})^{\alpha}\cdot \mathcal{H}_1(|\mathcal{U}|+1)^r,
      \qquad
      sk_3=g_2^r,
      \qquad
      sk_4=\mathcal{H}_1(ID_{SP})^r,
      \]
      
      where $|\mathcal{U}|$ denotes the number of the universe of attributes. 
      Output $sk_{SP}=(\mathcal{S},\{sk_{1,s}\}_{s\in \mathcal{S}},sk_2,sk_3,sk_4)$. \\
      \underline{GenSeed($mpk,ID_{PDO}$)}: Pick $\sigma \in \mathbb{Z}_p^*$, compute $\gamma =h(ID_{PDO}||\sigma)$ and $\psi=g_2^{\gamma}$. Output $\gamma$ and $\psi$. Send $\psi$ to TA. \\
      \underline{PKeyGenPDO($\psi$)}: With the input seed $\psi \in \mathbb{G}_2$, pick $\beta \in \mathbb{Z}_p$ and compute PDO's public key 
      $pk_{PDO}=\psi^{\beta}$. Output $pk_{PDO}$ and the factor $\beta$. Send $pk_{PDO},\beta$ to PDO. \\
      \underline{SKeyGenPDO($\gamma,\beta$)}: With the input seed $\gamma,\beta \in \mathbb{Z}_p$, set ${\gamma=\gamma \beta}$ and output $sk_{PDO}=\gamma$. \\
      \underline{Encapsulate($mpk,sk_{PDO},\mathcal{DG}_{n},\mathbb{A}$)}: 
      Initializing $c=1$ and the length of data granule $\ell$, for $k\in [c]$, pick $a_k \in \{0,1\}^{\ell}, d_k \in \mathbb{Z}_p^*$. 
      Compute $P_1=\oplus_{k=1}^{c}a_k, DCI=\prod_{k=1}^{c}g_2^{d_k}$. 
      Pick $y\in \mathbb{Z}_p$, compute $C_1=g_2^y$. Then, compute 
      \[
        P_2=\mathcal{H}_2(e(g_1^{sk_{PDO}},C_1)),
        \qquad
        C_2=(\oplus \{\mathcal{DG}_n\})\oplus P_1\oplus P_2,
        \]
      where $\oplus \{\mathcal{DG}_n\}$ denotes the XOR of all elements in the set $\mathcal{DG}_{n}$. 
      Parse $\mathbb{A}=(\mathbb{M},\pi)$. Suppose $\mathbb{M}$ has the shape $(n_1\times n_2)$. 
      Pick $\textbf{v} \in \mathbb{Z}_p^{n_2-1}, \textbf{y}' \in \mathbb{Z}_p^{\tau}$.
      For $j\in [\tau]$, compute $C_{3,j}=g_2^{\textbf{y}'[j]}$. For $i \in [n_1]$, compute 
      \[C_{4,i}=\mathcal{H}_1(|\mathcal{U}|+1)^{\mathbb{M}_i\cdot(y\Vert \textbf{v})^{\top}}
      \cdot \mathcal{H}_1(\pi(i))^{\textbf{y}'[\rho(i)]}\]
      Next, compute $\delta=\mathcal{H}_3(DCI,C_1,C_2,\{C_{3,j}\}_{j\in [\tau]},
      \{C_{4,i}\}_{i\in [n_1]})$ and $V=g_1^{\delta\sum_{k=1}^{c}d_k }$. Output $DCI$, 
      \[\mathcal{L}=(DCI,P_1=\bigoplus_{k=1}^{c}a_k,d=\sum_{k=1}^{c}d_k,y),
      \quad
      DC=(\mathbb{A},C_1,C_2,\{C_{3,j}\}_{j\in [\tau]},
      \{C_{4,i}\}_{i\in [n_1]},V).
        \]
    \underline{TaskIssue($mpk,sk_{PDO},ID_{SP},\mathcal{DG}_{n},\mathcal{I},\mathcal{L},t)$}: 
    $\mathcal{I}\subseteq [n]$ represents the indices of data granules that PDO wants to share with the SP. Parse $\mathcal{L}=(DCI,P_1,d,y)$.
    \begin{enumerate}
      \item{\textit{Task Generation Phase}}: Using $sk_{PDO}$, compute $\mathcal{T}_1=\mathcal{H}_1(ID_{SP})^{sk_{PDO}}\cdot \mathcal{H}_1(|\mathcal{U}|+1)^d$. 
      Using $g_2^{\alpha}$ from $mpk$, compute \[
        P_{\mathcal{T},1}=e(\mathcal{H}_1(ID_{SP})^d,g_2^{\alpha}),
        \qquad
        P_{\mathcal{T},2}=e(\mathcal{H}_1(ID_{SP})^y,g_2^{\alpha}),
        \qquad
        P_{\mathcal{T}}=P_{\mathcal{T},1}\cdot P_{\mathcal{T},2}.
        \]
      Compute $\mathcal{T}_2=P_{\mathcal{T}}\cdot e(g_1^{sk_{PDO}},g_2^y)$. 
      For $w\in \mathcal{I}$, pick $r_w\in \mathbb{Z}_p$ and compute 
      \[P_{w}=e(\mathcal{H}_1(ID_{SP})^{r_w},g_2^{\alpha}),
      \qquad
      \mathcal{T}_{w,1}=(\oplus\{\mathcal{DG}_{n}\backslash dg_w\})\oplus P_1\oplus \mathcal{H}_2(P_w),
      \qquad
      \mathcal{T}_{w,2}=P_{\mathcal{T}}\cdot P_{w}.
      \] 
      \item{\textit{Parameter Generation Phase}}: Pick $a_{c+1} \in \{0,1\}^{\ell}, d_{c+1} \in \mathbb{Z}_p^*$. Compute
      \[DCI'=DCI\cdot g_2^{d_{c+1}}=\prod_{k=1}^{c+1}g_2^{d_k},
      \qquad
      P_1'=P_1\oplus a_{c+1}=\bigoplus_{k=1}^{c+1}a_{k},
      \qquad
      d'=d+d_{c+1}=\sum_{k=1}^{c+1}d_{k}.\]
    \end{enumerate}
    Update $\mathcal{L}=(DCI',P_1',d',y)$. Compute $\mathcal{R}_1=g_1^{d'}=g_1^{\sum_{k=1}^{c+1}d_k}$. Set $\mathcal{T}_w=(\mathcal{T}_{w,1},\mathcal{T}_{w,2})$. Output $DCI$, 
      \[\mathcal{T}=(\mathcal{T}_1,\mathcal{T}_2,\mathcal{T}_3=\{\mathcal{T}_{w}\}_{w\in \mathcal{I}}),
      \qquad
      \mathcal{R}=(\mathcal{R}_1,\mathcal{R}_2=DCI',\mathcal{R}_3=a_{c+1}),
      \qquad
      \mathcal{D}=(\mathcal{D}_1=P_{\mathcal{T},1},\mathcal{D}_2=t).
        \]
        \underline{AccessDC$(mpk,sk_{SP},DCI,\mathcal{T},pk_{PDO})$}\footnote{
          Service providers run the AccessDC algorithm to check the authenticity of 
          the task issuer's identity, ensuring it corresponds to the expected 
          ID (\(pk_{PDO}\)). Upon successful verification, 
          the algorithm outputs a download parameter for further proceedings. 
          }: 
    Using $\mathcal{T}_1$ in $\mathcal{T}$ and $sk_2,sk_3,sk_4$ in $sk_{SP}$, compute and output
    \[P_{\mathcal{T},1}=\frac{e(sk_2,DCI)\cdot e(sk_4,pk_{PDO})}{e(\mathcal{T}_1,sk_3)}
          \]
        \underline{DownloadDC($DCI,\mathcal{D},P_{\mathcal{T},1}$)}: Determine 
        whether $t_{now}<\mathcal{D}_2$ and $P_{\mathcal{T},1}=\mathcal{D}_1$ holds, where $t_{now}$ denotes the current time. Return $\bot$ if not. 
        Otherwise, find the target $DC$ by $DCI$ and ouput $DC$. \\
  \end{tcolorbox}
  
  \caption{Our construction of TD-DCSS.}
  \label{fig:cons1}
\end{figure*}

\begin{figure*}[!htp]
  \begin{tcolorbox}[colback=white,colframe=black,arc=0pt,outer arc=0pt,width=\textwidth]
    \raggedright
    
    \underline{DecDC($mpk,sk_{SP},DCI,DC,\mathcal{T},P_{\mathcal{T},1}$)}: 
      \begin{enumerate}
        \item{\textit{DC Integrity Verification Phase.}}:  Parse $DC=(\mathbb{A},C_1,C_2,\{C_{3,j}\}_{j\in [\tau]},
        \{C_{4,i}\}_{i\in [n_1]},V)$. 
        Compute $\delta=\mathcal{H}_3(DCI,C_1,C_2,\{C_{3,j}\}_{j\in [\tau]},
        \{C_{4,i}\}_{i\in [n_1]})$. Return $\bot$ if $e(V,g_2)=e(g_1^{\delta},DCI)$ not holds. 
        \item{\textit{Attribute-based Decryption Phase}}: Compute $P_{\mathcal{T},2}=\ $FABEO.Dec$(mpk,\mathbb{A},\mathcal{S},CT_{FABEO},sk_{SP})$ if $\mathcal{S}$ in $sk_{SP}$ satisfies $\mathbb{A}$, 
        where $CT_{FABEO}=\{C_1,\{C_{3,j}\}_{j\in [\tau]},\{C_{4,i}\}_{i\in [n_1]}\}$. 
        {
            The detail of FABEO.Dec is shown as below: 
            If $\mathcal{S}$ in $sk_{SP}$ satisfies $\mathbb{A}$, there exists constants $\{\gamma_i\}_{i\in [I]} \ s.t. \sum_{i\in [I]}\gamma_i\mathbb{M}_i=(1,0,...,0)$, where $I=
            |\mathcal{S} \cap \{\pi(i)|i\in[n_1]\}|$.
            \[
              P_{\mathcal{T},2}= e(sk_2,C_1) \cdot \frac{\prod_{j\in [\tau]}e(\prod_{i\in [I],\rho(i)=j}(sk_{1,\pi(i)})^{\gamma_i},C_{3,j})}{e(\prod_{i\in [I]}(C_{4,i})^{\gamma_i},sk_3)}
            \]
            }
        Return $\bot$ if $P_{\mathcal{T},2}$ is invalid. 
        \item{\textit{Message Recovery Phase}}: Based on the results of above two phases, compute 
        $P_{\mathcal{T}}=P_{\mathcal{T},1}\cdot P_{\mathcal{T},2}$. Reconstruct $P_{2}$ by computing 
        \[\mathcal{H}_2(\frac{\mathcal{T}_2}{P_{\mathcal{T}}})=\mathcal{H}_2(\frac{e(\mathcal{H}_1(ID_{SP})^{\sum_{k=1}^cd_k},g_2^{\alpha})\cdot e(\mathcal{H}_1(ID_{SP})^y,g_2^{\alpha})\cdot e(g_1^{\gamma},g_2^y)}{e(\mathcal{H}_1(ID_{SP}),g_2)^{\alpha y}\cdot e(\mathcal{H}_1(ID_{SP})^{\alpha},\prod_{k=1}^{c}g_2^{d_k})})
          \]
        For $w\in \mathcal{I}$, compute 
        \[P_w=\frac{\mathcal{T}_{w,2}}{P_{\mathcal{T}}},
        \qquad
        dg_w = C_2\oplus \mathcal{T}_{w,1}\oplus \mathcal{H}_2(P_w)\oplus P_2. 
        \]
      \end{enumerate}
      Output $\{dg_w\}_{w\in \mathcal{I}}$. \\ 
      \underline{UpdateDC($mpk,DCI,DC,\mathcal{R}$)}\footnote{
        The UpdateDC algorithm modifies segments within data capsules. 
        Furthermore, upon updating, the new data capsule replaces 
        the old one, thus maintaining storage space without expansion.}: Update DC by setting $DCI'=\mathcal{R}_2$ and computing 
    \[C_2'=C_2\oplus \mathcal{R}_3,
    \qquad
    \delta'=\mathcal{H}_3(DCI',C_1,C_2',\{C_{3,j}\}_{j\in [\tau]},
      \{C_{4,i}\}_{i\in [n_1]}),
    \qquad
    V' = \mathcal{R}_1^{\delta'}.\]
    Output the updated $DCI=DCI'$, $DC=(\mathbb{A},C_1,C_2',\{C_{3,j}\}_{j\in [\tau]},
    \{C_{4,i}\}_{i\in [n_1]},V')$. 
  \end{tcolorbox}
  \caption{Our construction of TD-DCSS (Cont).}
  \label{fig:cons2}
  
\end{figure*}
\section{Proposed scheme}
\subsection{The Proposed TD-DCSS Scheme}
Our TD-DCSS scheme is shown in Fig. \ref{fig:cons1} and Fig. \ref{fig:cons2}. 
Specifically, when selectively sharing data with service providers, 
the personal data owner (PDO) first encapsulates its data into a data capsule, 
and then issues a task. 
Subsequently, the PDO sends the task to a service provider (SP) and its $PK_{PDO}$. 
The SP runs the AccessDC algorithm to check whether the $PK_{PDO}$ 
matches the public key of the task issuer and recover the download token. 
Then the SP successfully downloads the data capsule from the CS if 
the task is within its validity period. 
Finally, the SP can 
successfully recover the data if its attributes satisfy the access 
policy associated with the data capsule. In our scheme, 
the function FABEO.Dec serves as a tool that decrypts the ABE ciphertext, 
and our scheme is also designed to adopt other CP-ABE schemes that support useful 
features easily. 
\subsection{Correctness of Proposed TD-DCSS Scheme}
To decrypt a data capsule (or an updated data capsule generated by the UpdateDC algorithm) 
and recover the data granules, the SP first has to decrypt the task to 
obtain the download token $P_{\mathcal{T},1}$ through the AccessDC algorithm. 
Subsequently, the SP can recover the data granules $\{dg_w\}_{w\in\mathcal{I}}$ using 
the DecDC algorithm.  
The correctness of the AccessDC, DecDC, and UpdateDC algorithms is demonstrated below. \par
\textbf{Correctness of AccessDC}: $P_{\mathcal{T},1}^*$ can be computed in 
AccessDC algorithm as shown in 
Fig. \ref{fig:cons1}. \par
If the task is associated with the $ID_{SP}$ and the identifier $DCI$ of $DC$, 
then
\small{
  \begin{align*}
    P_{\mathcal{T},1}^*=&\frac{e(sk_2,DCI)\cdot e(sk_4,pk_{PDO})}{e(\mathcal{T}_1,sk_3)}\\
    &=\frac{e(sk_2,DCI)}{e(\mathcal{T}_1,sk_3)}\cdot e(sk_4,pk_{PDO})
\end{align*}}\par
For simplicity, we can divide the above equation into two parts: the left part $L$ and the right part $R$. 
We can compute 
  \begin{align*}
    L&=\frac{e(sk_2,DCI)}{e(\mathcal{T}_1,sk_3)}\\
    &=\frac{e(\mathcal{H}_1(ID_{SP})^{\alpha}\cdot \mathcal{H}_1(|\mathcal{U}|+1)^r,\prod_{k=1}^{c}g_2^{d_k})}{e(\mathcal{H}_1(ID_{SP})^{\gamma}\cdot \mathcal{H}_1(|\mathcal{U}|+1)^{\sum_{k=1}^{c}d_k},g_2^r)}\\
    &=\frac{e(\mathcal{H}_1(ID_{SP}),g_2)^{\alpha \sum_{k=1}^{c}d_k}\cdot e(\mathcal{H}_1(|\mathcal{U}|+1),g_2)^{r \sum_{k=1}^{c}d_k}}{e(\mathcal{H}_1(ID_{SP}),g_2)^{\gamma r}\cdot e(\mathcal{H}_1(|\mathcal{U}|+1),g_2)^{r \sum_{k=1}^{c}d_k}}\\
    &=\frac{e(\mathcal{H}_1(ID_{SP}),g_2)^{\alpha \sum_{k=1}^{c}d_k}}{e(\mathcal{H}_1(ID_{SP}),g_2)^{\gamma r}}\\
    \\
R&=e(sk_4,pk_{PDO})\\
\\
    L\cdot R&=\frac{e(\mathcal{H}_1(ID_{SP}),g_2)^{\alpha \sum_{k=1}^{c}d_k}}{e(\mathcal{H}_1(ID_{SP}),g_2)^{\gamma r}}\cdot e(sk_4,pk_{PDO})\\
    &=\frac{e(\mathcal{H}_1(ID_{SP}),g_2)^{\alpha \sum_{k=1}^{c}d_k}}{e(\mathcal{H}_1(ID_{SP}),g_2)^{\gamma r}}\cdot e(\mathcal{H}_1(ID_{SP}),g_2)^{\gamma r}\\
    &=e(\mathcal{H}_1(ID_{SP}),g_2)^{\alpha\cdot \sum_{k=1}^{c}d_k}=P_{\mathcal{T},1}\\
  \end{align*}\par
\textbf{Correctness of DecDC}: $\{dg_w\}_{w\in\mathcal{I}}^*$ can be computed in DecDC algorithm as shown in 
  Fig. \ref{fig:cons2}. In detail, $\delta^*$ can be computed as $\delta^*=\mathcal{H}_3(DCI,C_1,C_2,\{C_{3,j}\}_{j\in [\tau]},
  \{C_{4,i}\}_{i\in [n_1]})$ and the correctness of $\delta^*$ can be verified as 
  \[e(g_1^{\delta^*},DCI)=e(g_1,g_2)^{\delta^*\sum_{k=1}^{c}d_k}=e(g_1^{\delta^*\sum_{k=1}^{c}d_k},g_2)=e(V,g_2)\]\par
  The correctness of FABEO.Dec algorithm has been demonstrated\cite{riepel2022fabeo}, and it outputs the correct 
  $P_{\mathcal{T},2}$. 
  Based on the above results, we can compute 
  \begin{align*}
    \frac{\mathcal{T}_2}{P_{\mathcal{T}}}&=\frac{e(\mathcal{H}_1(ID_{SP})^{y+\sum_{k=1}^{c}d_k},g_2^{\alpha})\cdot e(g_1^{\gamma},g_2^y)}{P_{\mathcal{T},1}\cdot P_{\mathcal{T},2}}\\
    &=\frac{e(\mathcal{H}_1(ID_{SP})^{y+\sum_{k=1}^{c}d_k},g_2^{\alpha})\cdot e(g_1^{\gamma},g_2^y)}{e(\mathcal{H}_1(ID_{SP}),g_2)^{\alpha\cdot \sum_{k=1}^{c}d_k}\cdot e(\mathcal{H}_1(ID_{SP}),g_2)^{\alpha\cdot y}}\\
    &=e(g_1^{\gamma},g_2^y)
  \end{align*}\par
  We have $P_2^*=\mathcal{H}_2(\frac{\mathcal{T}_2}{P_{\mathcal{T}}})$. 
  For ${w\in\mathcal{I}}$, $dg_w^*$ can be computed as 
  \begin{align*}
    dg_w^*&=C_2\oplus \mathcal{T}_{w,1}\oplus \mathcal{H}_2(\frac{\mathcal{T}_{w,2}}{P_{\mathcal{T}}})\oplus P_2^*\\
    &=\oplus \{\mathcal{DG}_n\} \oplus P_1 \oplus P_2\oplus \mathcal{T}_{w,1}\oplus \mathcal{H}_2(\frac{\mathcal{T}_{w,2}}{P_{\mathcal{T}}})\oplus P_2^*\\
    &=\oplus \{\mathcal{DG}_n\} \oplus P_1 \oplus \{\mathcal{DG}_n\backslash dg_w\} \oplus P_1 \oplus \mathcal{H}_2(P_w) \oplus \mathcal{H}_2(P_w)^*\\
    &=dg_w
  \end{align*}\par
  \textbf{Correctness of UpdateDC}: The updated $DCI'$ can be computed as 
  $DCI'=g_2^{\sum_{k=1}^{c+1}d_k}=g_2^{d'}$. And the updated $DC'$ can be computed 
  by computing the updated $C_2'$ and $V'$. Specifically, the $C_2'$ can be computed 
  as $C_2'=C_2\oplus \mathcal{R}_3=(\oplus \{\mathcal{DG}_n\})\oplus P_1\oplus P_2\oplus a_{c+1}=(\oplus \{\mathcal{DG}_n\})\oplus P_1' \oplus P_2$. 
  And the $V'$ can be computed as 
  $V'=\mathcal{R}_1^{\delta'}=g_1^{\delta' \cdot\sum_{k=1}^{c+1}d_k}=g_1^{\delta'\cdot d'}$, 
  where $\delta'=\mathcal{H}_3(DCI',C_1,C_2',\{C_{3,j}\}_{j\in [\tau]},
  \{C_{4,i}\}_{i\in [n_1]})$. 
\subsection{Soundness of Proposed TD-DCSS Scheme}
\begin{table*}[!ht]
  \centering
  \caption{Property \& Functionality Comparisons}
  \label{tab:propertywise}
  \begin{threeparttable}
  \begin{tabularx}{\textwidth}{c|c|c|c|c|c|c|X}
  \toprule
  \textbf{Scheme} & \textbf{Selective Sharing} &\textbf{Informed-consent based Authorization} & \textbf{Permission Revocation} & \textbf{CR}& \textbf{TR} & \textbf{ER} & \centering \arraybackslash \textbf{Groups} \\
  \hline 
  DYL+\cite{dong2014achieving}&\ding{55}&\ding{55}&\ding{51}&\ding{51}&\ding{55}&\ding{55}&\centering \arraybackslash Symmetric\\
  \hline 
  ZSL+\cite{zuo2017fine}&\ding{55}&\ding{55}&\ding{51}&\ding{51}&\ding{55}&\ding{55}&\centering \arraybackslash Symmetric\\
  \hline 
  NHS+\cite{ning2020dual}&\ding{55}&\ding{55}&\ding{55}&\ding{51}&\textit{N/A}&\ding{51}&\centering \arraybackslash Symmetric\\
  \hline 
  YSX+\cite{yang2022efficient}&\ding{51}&\ding{51}&\ding{55}&\ding{55}&\ding{51}&\ding{55}&\centering \arraybackslash Symmetric\\
  \hline 
  TD-DCSS&\ding{51}&\ding{51}&\ding{51}&\ding{51}&\ding{51}&\ding{51}&\centering \arraybackslash Asymmetric\\
  \bottomrule
  \end{tabularx}
  \begin{tablenotes}
    \small
    \item \textbf{``CR''},\textbf{``TR''}, and \textbf{``ER''} are the abbreviation of \textbf{``Collusion Resistance''}, \textbf{``Tamper Resistance''}, and \textbf{``EDoS Resistance''}, respectively.
  \end{tablenotes}
\end{threeparttable}
\end{table*}
The soundness of our proposed TD-DCSS scheme encompasses three cases: 
soundness of AccessDC, soundness of DecDC, and no leakiness of UpdateDC. \par
For soundness of AccessDC, the PDO utilizes the secret key $sk_{PDO}$ to generate $\mathcal{T}_1$, 
and the PDO's public key takes the form $g_2^{sk_{PDO}}$. If the $sk_{PDO}$ used in 
$\mathcal{T}_1$ does not match $pk_{PDO} = g_2^{sk_{PDO}}$, the probability of 
recovering the correct $P_{\mathcal{T},1}$ is negligible. Furthermore, 
if an SP possesses an ``old'' or a ``wrong'' task and attempts to access the DC, 
the SP cannot successfully recover the correct $P_{\mathcal{T},1}$. 
In the first case, 
the $\mathcal{T}_1$ in the task is bound with the factor $d=\sum_{k=1}^{c}d_k$, 
the identifier $DCI$ is bound with the factor $d^* = \sum_{k=1}^{c+1}d_k$, and thus the 
components $e(\mathcal{H}_1(|\mathcal{U}|+1),g_2)^{r\cdot d}$ and 
$e(\mathcal{H}_1(|\mathcal{U}|+1),g_2)^{r\cdot d^*}$ cannot be canceled. 
In the second case, 
if an SP$^*$ holds a ``wrong'' task generated under the $ID_{SP}$, 
the SP$^*$ cannot recover a correct $P_{\mathcal{T},1}$ since 
$$
P_{\mathcal{T},1}^*=
\frac{e(\mathcal{H}_1(ID_{SP}^*)^{\alpha},g_2^d)\cdot e(\mathcal{H}_1(ID_{SP}^*)^r,g_2^{sk_{PDO}})}{e(\mathcal{H}_1(ID_{SP})^{sk_{PDO}},g_2^r)}
\neq P_{\mathcal{T},1}
$$
\par 
For soundness of DecDC, suppose an SP with an identifier $ID_{SP}$ holds a task $\mathcal{T}$, 
and another SP$^*$ with an identifier $ID_{SP^*}$ possesses sufficient attributes to 
satisfy the access policy associated with the DC. The SP first recovers 
$P_{\mathcal{T},1} = e(\mathcal{H}_1(ID_{SP}),g_2)^{\alpha\cdot d}$, 
and then the SP$^*$ recovers $P_{\mathcal{T},2}^* = 
e(\mathcal{H}_1(ID_{SP^*}),g_2)^{y\cdot \alpha}$. 
To recover 
data granules, they have to reconstruct $P_2$ by computing 
$$
P_2=\mathcal{H}_2(\frac{\mathcal{T}_2}{P_{\mathcal{T},1}\cdot P_{\mathcal{T},2}^*})=
\mathcal{H}_2(\frac{e(\mathcal{H}_1(ID_{SP})^y,g_2^{\alpha})\cdot e(g_1^{\gamma},g_2^y)}{e(\mathcal{H}_1(ID_{SP}^*),g_2)^{\alpha y}})
$$
The probability of the SP using $P_{\mathcal{T},2}^*$ to 
successfully recover $P_2$ is negligible, 
as $P_{\mathcal{T},2} = e(\mathcal{H}_1(ID_{SP}),g_2)^{y\cdot \alpha} \neq 
P_{\mathcal{T},2}^*$. 
The soundness of DecDC indicates that the 
proposed scheme is resistant to collusion attacks. 
\par
For no leakiness of UpdateDC, let's consider a scenario where the CS collects information during the 
update operation. The CS possesses information $a_2, a_3, ..., a_c$ but 
lacks any knowledge about $a_1$ and cannot recover $P_1$ with a non-negligible probability. 
Similarly, the CS is aware of $g_2^{d_1}, g_2^{d_2}, ..., g_2^{d_k}, 
g_1^{\sum_{k=2}^{c}d_k}$ but lacks any knowledge about $g_1^{d_1}$ and 
therefore cannot successfully recover $\sum_{k=1}^{c}d_k$ with a non-negligible 
probability.
\subsection{Security Analysis of Proposed TD-DCSS Scheme}
\textbf{\textit{Theorem 1}}: Assume that DBDH assumption holds, then 
the proposed $\mathcal{DCSS}_{\mathcal{TD}}$ is 
$\mathsf{IND}$-$\mathsf{CPA}$ secure. \par
\textbf{\textit{Proof}}: If a polynomial time adversary $\mathcal{A}$ is capable of 
successfully breaching the TD-DCSS, then by interacting with $\mathcal{A}$, 
another algorithm $\mathcal{C}$ can be easily constructed to exploit the DBDH 
assumption and break the $\mathsf{IND}$-$\mathsf{CPA}$ security of the TD-DCSS. 
Given a DBDH tuple $(g^a, g^b, g^c, h^a, h^b, h^c, \mathcal{Z})$ to $\mathcal{C}$, where 
$g\in \mathbb{G}_1,h\in \mathbb{G}_2$, 
the goal of $\mathcal{A}$ is to determine whether $\mathcal{Z} = e(g, h)^{abc}$ 
or $\mathcal{Z} = \mathcal{R}$, where $\mathcal{R}$ is a random 
element of $\mathbb{G}_T$.\par
\textbf{Init.} $\mathcal{A}$ submits a challenge access 
policy $\mathbb{A}^*=(\mathbb{M}^*,\pi^*)$ 
to $\mathcal{C}$.\par
\textbf{Setup.} $\mathcal{C}$ runs GroupGen($1^{\lambda}$) algorithm to 
generate ($p,\mathbb{G}_1,\mathbb{G}_2,\mathbb{G}_T,e,g_1,g_2$). Then 
$\mathcal{C}$ sets $g_1=g^a$ and $g_2=h^b$. Pick $\alpha\in \mathbb{Z}_p^*$, 
compute $g_2^{\alpha}=h^{b\alpha}$. $\mathcal{C}$ sends the 
$mpk=(p,\mathbb{G}_1,\mathbb{G}_2,\mathbb{G}_T,e,g^a,h^b,h^{b\alpha},h,\mathcal{H}_{i\in[3]})$ to $\mathcal{C}$, 
where $h,\mathcal{H}_{i\in[3]}$ are collision-resistant hash function. The 
master private key is set as $msk=(\alpha)$.\par
\textbf{Phase 1.} $\mathcal{A}$ can adaptively send a sequence of the following 
queries to $\mathcal{C}$.
\begin{itemize}
  \item{
    $\mathcal{O}_{\text{KSP}}(ID_{SP},\mathcal{S}\subseteq \mathcal{U})$: On input 
    $ID_{SP}\in \{0,1\}^*$ and $\mathcal{S}\subseteq \mathcal{U}$, $\mathcal{C}$ checks 
    whether $\mathcal{S}\models \mathbb{A}^*$ holds. If $\mathcal{S}\models \mathbb{A}^*$ 
    holds, $\mathcal{C}$ aborts and returns $\bot$. Otherwise, it 
    picks $r\in \mathbb{Z}_p^*$ and 
    sets 
    $sk_{1,s}=\mathcal{H}_1(s)^r,
    sk_2=\mathcal{H}_1(ID_{SP})^{\alpha}\cdot \mathcal{H}_1(|\mathcal{U}|+1)^r,
    sk_3=h^{br},
    sk_4=\mathcal{H}_1(ID_{SP})^r
    $, where $s\in \mathcal{S}$.
  }
  \item{
    $\mathcal{O}_{\text{GenSeed}}(ID_{PDO})$: On input 
    $ID_{PDO}\in \{0,1\}^*$, $\mathcal{C}$ picks $\sigma\in\mathbb{Z}_p^*$ and 
    computes $\gamma=h(ID_{PDO}||\sigma)$. Then $\mathcal{C}$ 
    sets $\psi=h^{b\gamma}$. $\mathcal{C}$ inits a table 
    $D_1$ and records the record $(ID_{PDO},\gamma)$ into $D_1$. And then 
    $\mathcal{C}$ sends the $\psi$ to $\mathcal{A}$. 
  }
  \item{
    $\mathcal{O}_{\text{PKPDO}}(\psi)$: On input 
    $\psi \in \mathbb{G}_2$, $\mathcal{C}$ picks $\beta\in \mathbb{Z}_p^*$ and sets 
    $pk_{PDO}=\psi^{\beta}=h^{b\gamma \beta}$. Then $\mathcal{C}$ sends 
    $\beta,pk_{PDO}$ to $\mathcal{A}$.  
  }
  \item{
    $\mathcal{O}_{\text{SKPDO}}(ID_{PDO},\beta)$: On input $ID_{PDO} \in \{0,1\}^*$ and 
    $\beta\in \mathbb{Z}_p^*$, $\mathcal{C}$ first searches table $D_1$ to 
    retrieve $\gamma$. If there does not exist a record $(ID_{PDO},\gamma)$ in $D_1$, it return  
    $\bot$. Otherwise, $\mathcal{C}$ sets $sk_{PDO}=\gamma\cdot \beta$. 
    After that, 
    it sends $sk_{PDO}$ to $\mathcal{A}$. 
  }
\end{itemize}

Note that the seed generation oracle $\mathcal{O}_{\text{GenSeed}}(ID_{PDO})$, 
the public key generation oracle $\mathcal{O}_{\text{PKPDO}}(\psi)$, 
and the secret generation oracle $\mathcal{O}_{\text{SKPDO}}(\gamma,\beta)$ 
must be queried in a sequential order. \par
\textbf{Challenge.} $\mathcal{A}$ submits two sets of data granules $(m_0,m_1)$ with 
the same length $\ell$ to $\mathcal{C}$. $\mathcal{C}$ first picks 
$P_1\in \{0,1\}^{\ell}$ and $d,y\in \mathbb{Z}_p^*$, and computes 
$DCI^*=h^{bd}$ and $C_1=h^{by}$. 
Then, $\mathcal{C}$ sets $P_2=\mathcal{H}_2(\mathcal{Z}^{y})$.
$\mathcal{C}$ flips a random coin $b\in\{0,1\}$ and sets 
$C_2=m_b\oplus P_1 \oplus P_2$. Parse $\mathbb{A}^*=(\mathbb{M}^*,\pi^*)$. 
Suppose $\mathbb{M}^*$ has the shape $(n_1\times n_2)$. 
$\mathcal{C}$ picks $\textbf{v} \in \mathbb{Z}_p^{n_2-1}, \textbf{y}' \in \mathbb{Z}_p^{\tau}$.
For $j\in [\tau]$, it computes $C_{3,j}=h^{b\textbf{y}'[j]}$. 
For $i \in [n_1]$, it computes 
$C_{4,i}=\mathcal{H}_1(|\mathcal{U}|+1)^{\mathbb{M}_i\cdot(y\Vert \textbf{v})^{\top}}
\cdot \mathcal{H}_1(\pi(i))^{\textbf{y}'[\rho(i)]}$.
Next, it computes $\delta=\mathcal{H}_3(DCI^*,C_1,C_2,\{C_{3,j}\}_{j\in [\tau]},
\{C_{4,i}\}_{i\in [n_1]})$ and $V=g^{a\delta\sum_{k=1}^{c}d_k }$. 
Then $\mathcal{C}$ sends $DCI^*$ and 
$DC^*=(\mathbb{A^*},C_1,C_2,\{C_{3,j}\}_{j\in [\tau]},
\{C_{4,i}\}_{i\in [n_1]},V)$ to $\mathcal{A}$.\par
\textbf{Phase 2}. $\mathcal{A}$ continues to query $\mathcal{C}$ under the 
previous restrictions. \par
\textbf{Guess}. $\mathcal{A}$ outputs a guess $b'$. It returns 1 implying 
$\mathcal{Z}=e(g,h)^{abc}$ if $b'=b$. Otherwise, it returns 0 implying 
$\mathcal{Z}=\mathcal{R}$. \par
If $\mathcal{Z}=e(g,h)^{abc}$, then $sk_{PDO}^*=c$, where $sk_{PDO}^*$ is the secret key that used 
in the encryption, the challenge ciphertext queried by $\mathcal{A}$ originates from 
a distribution that is the same as in the construction. If $\mathcal{Z}=\mathcal{R}$, 
and since $\mathcal{R}$ is a random element of $\mathbb{G}_T$, the challenge ciphertext 
$C_2=m_b\oplus \mathcal{H}_2(\mathcal{R}^y) \oplus P_2$ is uniformly random. $m_b$ is 
independent in the view of $\mathcal{A}$. \par
In this simulation, the advantage 
of $\mathcal{A}$ 
is given by
\begin{align*}
  &\mathbf{Adv}^{\mathsf{IND}\text{-}\mathsf{CPA}}_{\mathcal{DCSS}_{\mathcal{TD}},\mathcal{A}}(\lambda)=\frac{1}{2}\cdot \frac{1}{2} + \frac{1}{2}\cdot(\frac{1}{2}+\epsilon)-\frac{1}{2}=\frac{\epsilon}{2}
\end{align*}
\hspace{0.2pt}
\section{Performance Evaluation}
In this section, many state-of-the-art schemes \cite{yang2022efficient},\cite{dong2014achieving},\cite{zuo2017fine},\cite{ning2020dual} 
are compared with our scheme. Firstly, we analyze these several related schemes regarding 
properties and functionality. Then, we give comprehensive computation 
and storage cost comparisons and an experimental simulation for our scheme to 
demonstrate its 
practicality.
\subsection{Property \& Functionality}
As shown in TABLE \ref{tab:propertywise}, several ABE-based data sharing schemes are 
compared with TD-DCSS of properties and functionality. 
We use \ding{51}(\ding{55}) to denote that the scheme achieves (not achieves) this property or 
functionality. ``N/A'' means this property or 
functionality is not applicable in this scheme. For simplicity, the 
CR, TR, and ER represent collusion resistance, tamper resistance, and 
EDoS resistance, respectively. 
\par
Many data sharing schemes based on ABE\cite{dong2014achieving},\cite{zuo2017fine},\cite{ning2020dual},\cite{yang2022efficient} 
have been proposed to realize fine-grained access control and privacy protection in data sharing. 
DYL+\cite{dong2014achieving} applies ABE and IBE to design a data sharing system 
that supports permission revocation. Furthermore, 
ZSL+\cite{zuo2017fine} introduces a two-factor mechanism to protect the 
security of users' secret keys. It can revoke secret keys efficiently based on the 
proxy re-encryption\cite{blaze1998divertible} and key 
separation techniques. 
Moreover, when considering cloud-hosted data, the economics of data sharing is important. 
To resist EDoS attacks, a dual access control is proposed by NHS+\cite{ning2020dual}. 
It achieves efficient access control over the download requests and protects 
sensitive data. 
To realize selective sharing and informed-consent based authorization, 
YSX+\cite{yang2022efficient} introduces a selective data sharing system based on ABE and SE. 
However, it neither realizes permission revocation nor resists collusion or EDoS attacks. 
Our TD-DCSS realizes the personal data sharing autonomy, including selective sharing, 
informed-consent based authorization, and permission revocation, with 
collusion, tamper, and EDoS resistance. 
Furthermore, since symmetric groups have serious security issues\cite{FAME}, our proposed TD-DCSS scheme is 
based on the asymmetric prime-order groups which support efficient hashing to $\mathbb{G}_1$\cite{riepel2022fabeo},\cite{rfc9380}. 
\par
In conclusion, our proposed TD-DCSS scheme has desirable properties and functionality 
superior to the state-of-the-art solutions. We give theoretical analysis via computation 
and storage cost comparisons to evaluate the efficiency of our scheme in the following. 
{
\begin{table*}[!htbp]
    \centering
    \begin{threeparttable}
    \caption{Computation cost comparisons}
    \label{tab:Computation complexity}
    \renewcommand{\arraystretch}{1}
    \begin{tabularx}{\textwidth}{p{0.075\textwidth}|c|p{0.05\textwidth}|c|X|c|X|X}
      \toprule
      \centering \textbf{Scheme} & \textbf{Setup} & \textbf{KGDO} & \centering\textbf{KGDU} & \centering \textbf{Enc} &\centering \textbf{PreWork} &\centering\textbf{Dec} & \centering \arraybackslash \textbf{Revoke} \\
      \hline
      \centering DYL+\cite{dong2014achieving}
      & \centering $|\mathcal{U}|(e_1+e_T)+p$
      & \centering 0 
      & \centering $2|\mathcal{S}|e_1$
      &  $3n_1e_1+(2n_1+1)e_T$
      & \centering -
      & \centering $2Ip$
      & \centering \arraybackslash $2n_1e_T$\\
      \hline
      \centering ZSL+\cite{zuo2017fine}
      & \centering $3|\mathcal{U}|e_1+e_T+p$
      & \centering 0 
      & \centering $(3|\mathcal{U}|+1)e_1$
      &  $(|\mathcal{U}|+1)e_1+e_T$
      & \centering CS:$(|\mathcal{U}|+1)e_1+e_T$
      & \centering $(|\mathcal{U}|+1)p$
      & \centering \arraybackslash $3|\mathcal{U}|e_1$\\
      \hline
      \centering NHS+\cite{ning2020dual}
      & \centering $e_1+e_T+p$
      & \centering 0 
      & \centering $(|\mathcal{S}|+2)e_1$
      & $(3n_1+1)e_1+e_T$
      & \centering
        \begin{tabular}{@{}c@{}}
          
          DU:$(|\mathcal{S}|+2)e_1$ \\
          Other:$e_1+(2I+1)p$ \\
          S:$(|\mathcal{S}|+3)e_1+(2I+1)p$
        \end{tabular}
      & \centering $(2I+1)p$
      & \centering \arraybackslash -\\
      \hline
      \centering YSX+\cite{yang2022efficient}
      & \centering$4e_1+e_T+p$ 
      & \centering0 
      & \centering$(|\mathcal{S}|+7)e_1$
      & $(3n_1+5)e_1+2e_T+p$ 
      & DU:$(|\mathcal{S}|+12)e_1$ 
      & \begin{tabular}{@{}X@{}}
        \centering
          CS:$(2I+7)p$ \\
          DU:$e_1+2e_T$ \\
          S:$e_1+e_T+(2I+7)p$
        \end{tabular}
      & \centering \arraybackslash -\\
      \hline
      \centering TD-DCSS
      & \centering$e_2$ 
      & \begin{tabular}{@{}c@{}}
          DO:$e_2$ \\
          TA:$e_2$ \\
          S:$2e_2$
        \end{tabular}  
      & \centering$(|\mathcal{S}|+3)e_1+e_2$
      & $(2n_1+2)e_1+(\tau +2)e_2+p$ 
      & DO:$(n+1)e_1+(n+2)p$ 
      & $e_1+(\tau +4)p$
      & \centering \arraybackslash 
        \begin{tabular}{@{}X@{}}
          \centering
          DO:$e_1+e_2$\\
          CS:$e_1$\\
          S:$2e_1+e_2$
        \end{tabular}\\
      \bottomrule
    \end{tabularx}
    \begin{tablenotes}
      \small
      \item ``\textbf{KGDO}''/``\textbf{KGDU}'' denote key generation for a data owner/data user; ``\textbf{PreWork}'' denotes the preparatory work for data sharing.
    \end{tablenotes}
  \end{threeparttable}
  \end{table*}
}
  \begin{table*}[!htbp]
    \centering
  \begin{threeparttable}
    \caption{Storage cost comparisons}
    \label{tab:Storage complexity}
    \renewcommand{\arraystretch}{1}
    \begin{tabularx}{\textwidth}{c|c|c|c|c|X}
      \toprule
      \textbf{Scheme} & \textbf{MPK} & \textbf{KeyDO} & \textbf{KeyDU} & \textbf{CT} & \centering \arraybackslash  \textbf{PreWork Cost}  \\
      \hline
      DYL+\cite{dong2014achieving}
      & \centering $|\mathcal{U}|(|\mathbb{G}_1|+|\mathbb{G}_T|)$ 
      & $0$
      & $|\mathcal{S}||\mathbb{G}_1|$ 
      & $|\mathbb{A}|+n_1(|\mathbb{G}_1|+|\mathbb{G}_T|)$
      & \centering \arraybackslash - \\
      \hline
      ZSL+\cite{zuo2017fine}
      & \centering $|\mathcal{U}||\mathbb{G}_1| + |\mathbb{G}_T|$ 
      & $0$
      & $(3|\mathcal{U}|+1)|\mathbb{G}_1|$ 
      & $|\mathbb{A}|+|\mathbb{G}_T|+(|\mathcal{U}|+1)|\mathbb{G}_1|$
      & \centering \arraybackslash - \\
      \hline
      NHS+\cite{ning2020dual}
      & \centering $|\mathbb{G}_1|+|\mathbb{G}_T|$ 
      & $0$
      & $(|\mathcal{S}|+2)|\mathbb{G}_1|$ 
      & $|\mathbb{A}|+|\mathbb{G}_T|+(3n_1+1)|\mathbb{G}_1|$
      & \centering \arraybackslash $(|\mathcal{S}|+2)|\mathbb{G}_1|$ \\
      \hline
      YSX+ \cite{yang2022efficient}
      & $4|\mathbb{G}_1|+|\mathbb{G}_T|$ 
      & $0$
      & $2|\mathbb{Z}_p|+(|\mathcal{S}|+4)|\mathbb{G}_1|$ 
      & $|\mathbb{A}|+3\ell+(2n_1+5)|\mathbb{G}_1|$
      & \centering \arraybackslash $(|\mathcal{S}|+8)|\mathbb{G}_1|+|\mathbb{Z}_p|$ \\
      \hline
      TD-DCSS
      & $|\mathbb{G}_2|$ 
      & $|\mathbb{Z}_p|$
      & $(|\mathcal{S}|+2)|\mathbb{G}_1|+|\mathbb{G}_2|$ 
      & $|\mathbb{A}|+\ell+(n_1+2)|\mathbb{G}_1|+(\tau+1)|\mathbb{G}_2|$
      & \centering \arraybackslash $n\ell+|\mathbb{G}_1|+(n+1)|\mathbb{G}_T|$ \\
      \bottomrule
    \end{tabularx}
    \begin{tablenotes}
      \small
      \item ``\textbf{MPK}'' and ``\textbf{CT}'' denote ``Master Public Key'' and ``Ciphertext'', respectively; ``\textbf{KeyDO}''/``\textbf{KeyDU}'' denote the storage consumption of the key of a data owner/a data user, respectively; ``\textbf{PreWork Cost}'' denote the storage cost of the preparatory work for data sharing.
    \end{tablenotes}
  \end{threeparttable}
  \end{table*}
\subsection{Computation \& Storage Cost}
As shown in TABLE \ref{tab:Computation complexity} and TABLE \ref{tab:Storage complexity}, we 
analyze those state-of-the-art solutions in terms of computation and storage overheads. 
In our comparisons, we are mainly considering the most time-consuming operations such as 
exponentiation and bilinear pairings. To show the results intuitively, we let 
$n,n_1$ be the number of the data granules that a PDO wants to share with an SP in the TaskIssue 
algorithm and the number of the rows of policy matrix $\mathbb{M}$, respectively. 
Let $\mathcal{|U|},\mathcal{|S|},I$ denote the number of the universe of attributes, the number 
of the attribute set of the data owner, and the number of attributes used in the decryption of the ABE scheme. 
In TABLE \ref{tab:Computation complexity}, we let $e_1,e_2,e_T,p$ be the overhead of 
a single exponentiation computation in $\mathbb{G}_1$, 
a single exponentiation computation in $\mathbb{G}_2$, 
a single exponentiation computation in $\mathbb{G}_T$ 
and a pairing operation, respectively. 
In TABLE \ref{tab:Storage complexity}, we let $|\mathbb{Z}_p|,|\mathbb{G}_1|,|\mathbb{G}_2|,|\mathbb{G}_T|$ be the size of 
a single element in $\mathbb{Z}_p,\mathbb{G}_1,\mathbb{G}_2$ and $\mathbb{G}_T$, respectively. 
Let $|\ell|,|\mathbb{A}|$ denote the length of a single data granule and 
the size of an access policy, respectively. \par
As shown in TABLE \ref{tab:Computation complexity}, we analyze the several phases: 
\textbf{Setup, KGDO, KGDU, Enc, PreWork, Dec, Revoke}. 
In detail, 
\textbf{KGDO, KGDU} means generating a secret key for data owners and generating a secret key for data users, 
respectively. The phase \textbf{PreWork} means the time cost of the preparatory work for 
data sharing, such as TaskIssue algorithm in TD-DCSS, 
data download in ZSL+\cite{zuo2017fine},NHS+\cite{ning2020dual} and 
keyword test in YSX+\cite{yang2022efficient}. To better demonstrate the computation overhead borne by different sides at the same stage, 
we let the prefix indicate that the time cost is borne by this party. Especially, the prefix 
``S:'' denotes the sum of all sides. \par
We can easily conclude that the computation cost of the \textbf{Setup} in NHS+\cite{ning2020dual}, 
YSX+\cite{yang2022efficient} and TD-DCSS is constant, while the 
cost of DYL+\cite{dong2014achieving} and ZSL+\cite{zuo2017fine} is growing 
with the $|\mathcal{U}|$ linearly. Our scheme has an additional phase \textbf{KGDO} 
but the cost is constant. 
The cost of \textbf{KGDU}, \textbf{Enc} in ZSL+\cite{zuo2017fine} follows a linear relationship with 
the $|\mathcal{U}|$ while other schemes follow a linear relationship with 
the $|\mathcal{S}|$ (in \textbf{KGDU}) or $n_1$ (in \textbf{Enc}) ($|\mathcal{S}|\leq |\mathcal{U}|$). Note that $\tau\leq I$ and we consider 
$\tau=1$. We can observe that the cost of \textbf{PreWork} in TD-DCSS is growing 
with $n$. The computation overhead of \textbf{PreWork} in TD-DCSS is constant 
if $n=1$, while other schemes 
follow a linear relationship with $|\mathcal{U}|$ or $|\mathcal{S}|$. We can also 
demonstrate that the cost of TD-DCSS is constant in \textbf{Dec} since we consider $\tau=1$, while other 
schemes are growing with $I$ or $|\mathcal{U}|$, where the cost at data owner side 
is constant in YSX+\cite{yang2022efficient}. 
The computation cost of \textbf{Revoke} in TD-DCSS is constant, while 
in DYL+\cite{dong2014achieving} and ZSL+\cite{zuo2017fine} is linear. \par
As depicted in TABLE \ref{tab:Storage complexity}, we mainly consider the most storage-consuming 
parts, such as master public key (\textbf{MPK}), secret key of data owner (\textbf{KeyDO}), 
secret key of data user (\textbf{KeyDU}), ciphertext (\textbf{CT}) and preparatory work cost 
(\textbf{PreWork}). We can find that the storage cost of \textbf{MPK} in 
NHS+\cite{ning2020dual},YSX+\cite{yang2022efficient} and TD-DCSS is constant while it in 
DYL+\cite{dong2014achieving} and ZSL+\cite{zuo2017fine} is growing with $|\mathcal{U}|$. 
The storage cost of \textbf{KeyDO} in TD-DCSS is constant. It is easy to observe that 
the storage cost of \textbf{KeyDU}, \textbf{CT} in all schemes except for ZSL+\cite{zuo2017fine}, has a linear relationship with 
$|\mathcal{S}|$ (\textbf{KeyDU}) and $n_1$ (\textbf{CT}), while ZSL+\cite{zuo2017fine} has a linear relationship with 
$|\mathcal{U}|$ of \textbf{KeyDU} and \textbf{CT}. When $n=1$, the overhead of 
\textbf{PreWork} in TD-DCSS is constant while it in \cite{ning2020dual} and \cite{yang2022efficient} 
is growing with $|\mathcal{S}|$. \par
\subsection{Experimental Analysis}
  \begin{figure*}[!ht]
    \centering
    \includegraphics[width=0.83\textwidth]{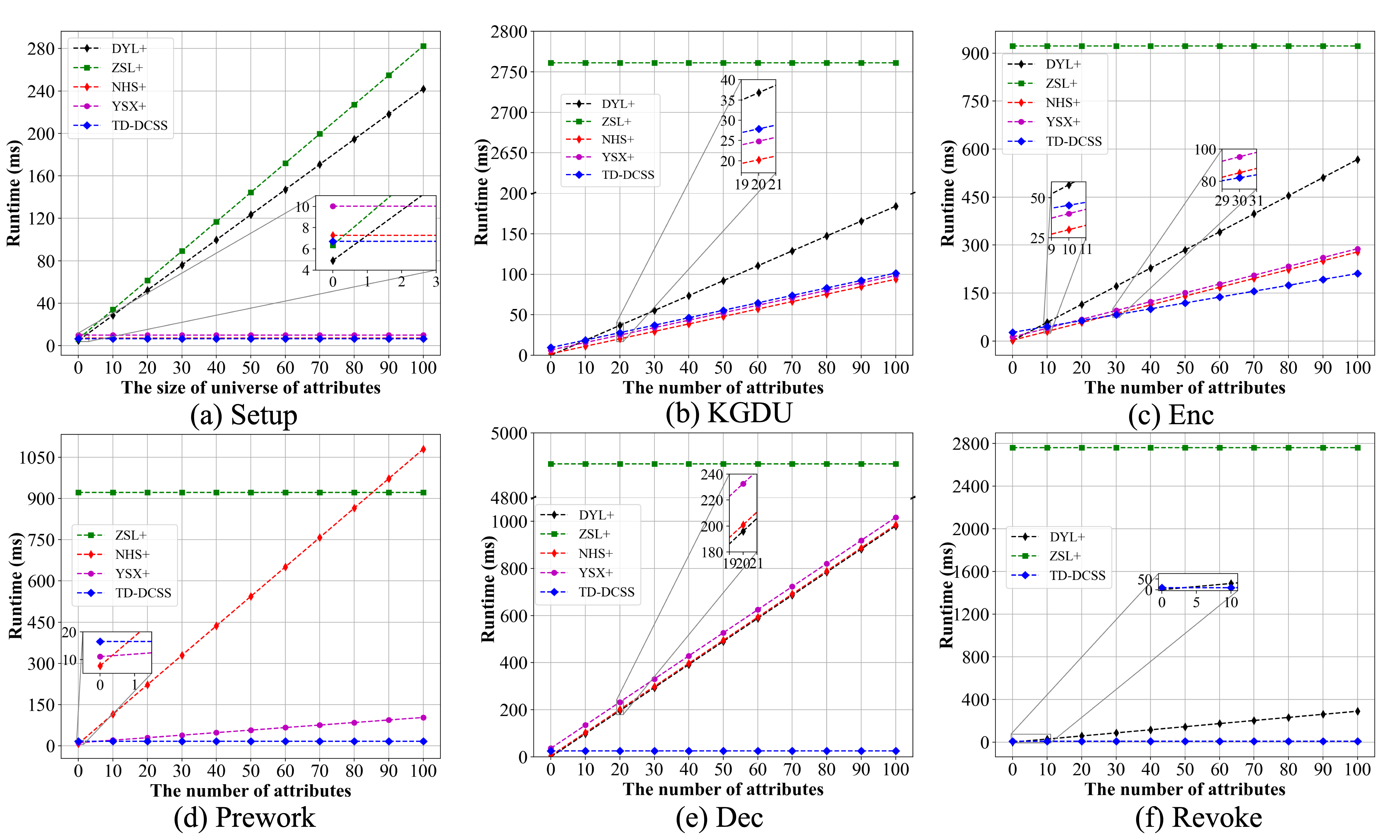}
    \caption{Runtime evaluation of Setup, KGDU, Enc, PreWork, Dec and Revoke phases. }
    \label{fig:performance}
  \end{figure*}
  \begin{figure*}[!t]
    \centering
    \includegraphics[width=0.95\textwidth]{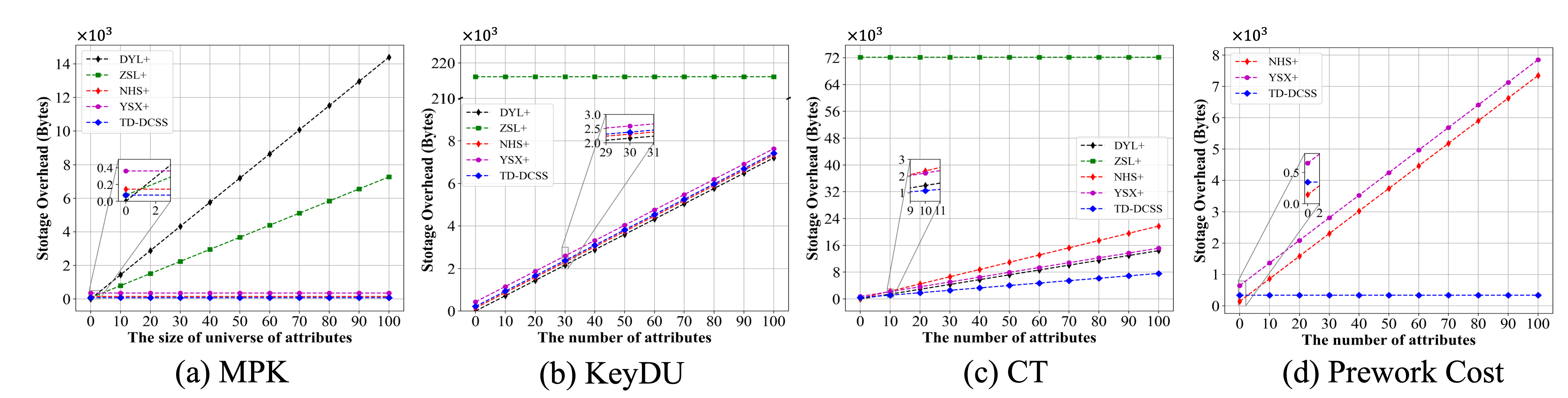}
    \caption{Storage overhead of MPK, KeyDU, CT and PreWork Cost.}
    \label{fig:storage per}
  \end{figure*}
  In order to demonstrate the practicality of our proposed scheme and the 
  effectiveness of the task mechanism in protecting data privacy, 
  we conducted an experimental simulation for our TD-DCSS scheme and other 
  schemes, 
  assessing their performance across various conditions. 
  The experiments were conducted on a standard personal computer running Windows 10, 
  equipped with an Intel(R) Core(TM) i7-7700HQ CPU @ 2.80GHz, 
  and 16GB of RAM. Our code, implemented in Python 3.6.9, 
  relies on the Charm 0.50 library \cite{akinyele2013charm}, 
  PBC-0.5.14 library, and operates within the Ubuntu 18.04.7 LTS environment (WSL).
  Our implementation based on the project available at 
  https://github.com/DoreenRiepel/FABEO, adopting $\ell=128$ and 
  utilizing the MNT224 curve. The source code for our TD-DCSS, 
  along with the corresponding evaluation code, 
  has been made publicly accessible at https://github.com/xiahezzz/TD-DCSS. 
  Since the fact that 
  $\mathcal{S}\subseteq\mathcal{U}$ and $\mathcal{|S|}<<\mathcal{|U|}$, we let 
  $\mathcal{|U|}$ be equal to 1000. 
  The outcomes of our experimental analysis are illustrated in Fig. \ref{fig:performance} and Fig. \ref{fig:storage per}.\par
  Fig. \ref{fig:performance} demonstrates the overhead of the 
  \textbf{Setup, KGDU, Enc, Prework, Dec and Revoke} phases. As shown in Fig. \ref{fig:performance}{\textcolor[rgb]{0,0,1}a}, 
  it is evident that the runtime of \textbf{Setup} in 
  NHS+\cite{ning2020dual} and YSX+\cite{yang2022efficient}. 
  In Fig. \ref{fig:performance}{\textcolor[rgb]{0,0,1}b} and 
  Fig. \ref{fig:performance}{\textcolor[rgb]{0,0,1}c}, 
  the runtime of \textbf{KGDU} and \textbf{Enc} of our TD-DCSS grows 
  as the number of attributes in $\mathcal{S}$ and $\mathbb{A}$ grows, respectively. 
  Moreover, the runtime of \textbf{Enc} in our TD-DCSS has a relatively constant 
  and lower computational overhead compared to other schemes. 
  Meanwhile, the runtime of the \textbf{KGDU} in our TD-DCSS consumes only a 
  slightly more constant time than NHS+\cite{ning2020dual}, YSX+\cite{yang2022efficient}. 
  Fig. \ref{fig:performance}{\textcolor[rgb]{0,0,1}d} shows the relationship 
  between the runtime of the \textbf{PreWork} and the number of attributes 
  in $\mathcal{S}$ (or the number of data granules in the task). 
  For better comparisons, we set $|\mathcal{I}| = |\mathcal{S}|$, and $n=1$ in 
  TD-DCSS, as other schemes share one data granule. 
  The results show that when $n=1$, the runtime of the \textbf{PreWork} in our 
  TD-DCSS remains constant, while in other schemes, it is linear. 
  As shown in Fig. \ref{fig:performance}{\textcolor[rgb]{0,0,1}e} and Fig. \ref{fig:performance}{\textcolor[rgb]{0,0,1}f}, 
  the runtime of the \textbf{Dec} and \textbf{Revoke} phases in our TD-DCSS is constant, taking 
  relatively less computation overhead compared to other schemes. The efficient execution of the UpdateDC algorithm can be attributed to 
  the parameters generated by the PDO, allowing the CS to readily update the data 
  capsule without requiring an update to the ABE-related ciphertext. \par
In our simulation, the runtimes for the GenSeed, PKeyGenPDO, and SKeyGenPDO 
algorithms in our TD-DCSS are 7ms, 3.7e-04ms, and 6ms, respectively. \par
Fig. \ref{fig:storage per} depicts the storage overhead for \textbf{MPK, KeyDU, CT} 
and \textbf{PreWork Cost}. 
From Fig. \ref{fig:storage per}{\textcolor[rgb]{0,0,1}a}, 
we can find that the storage overhead of \textbf{MPK} in NHS+\cite{ning2020dual}, 
YSX+\cite{yang2022efficient}, 
and our TD-DCSS is constant, 
while in DYL+\cite{dong2014achieving} and ZSL+\cite{zuo2017fine}, it is linear with the $\mathcal{|U|}$.  
As shown in Fig. \ref{fig:storage per}{\textcolor[rgb]{0,0,1}b}, 
the storage overhead of \textbf{KeyDU} in all works except ZSL+\cite{zuo2017fine} is linear 
with respect to $\mathcal{S}$. 
From Fig. \ref{fig:storage per}{\textcolor[rgb]{0,0,1}c}, it is 
easy to find that the storage overhead of \textbf{CT} in all works except for ZSL+\cite{zuo2017fine} is linear 
with the number of attributes in $\mathbb{A}$, and the \textbf{CT} in our TD-DCSS 
consumes less storage overhead. In Fig. \ref{fig:storage per}{\textcolor[rgb]{0,0,1}d}, 
we set $n=1$, as other schemes also share one data granule, and then we can find that 
the storage overhead of \textbf{PreWork Cost} in our TD-DCSS remains constant, while in other schemes, 
it is linear with the $\mathcal{|S|}$. 
\par  
These results demonstrate that the TD-DCSS scheme realizes personal data sharing 
for protecting owners' privacy and enabling fine-grained data sharing, with
acceptable runtime and storage consumption. 
\section{Conslusion and future work}
In this paper, we first introduce the concept of personal 
data sharing autonomy, 
including selective sharing, informed-consent based authorization 
and permission revocation, focusing on data owners' privacy. 
To realize the personal data sharing autonomy, 
we then propose the task-driven data capsule data sharing system (TD-DCSS), which mainly includes 
the data capsule encapsulation method and the task-driven data sharing mechanism. 
Finally, a comprehensive security analysis, with a theoretical analysis 
and an experimental simulation are conducted. 
The result reveals that our scheme is 
correct, sound, and secure under the security model, 
and is proved to be an effective scheme compared with state-of-the-art schemes. 
\par
In our scheme, a task can be used to access the data capsule only once, 
which enables a data owner has the right of 
informed-consent based authorization and permission revocation, precisely. 
However, this may lead to inefficient data sharing in the case that 
the same data need to be accessed frequently. 
Therefore, a possible future direction could involve designing a new 
mechanism that allows one task to be used multiple times while keeping 
the informed-consent based authorization and permission revocation. 
\par
\section{Acknowledgements}
We would like to thank Professor Jian Weng, Jinan University, China, 
for generously providing valuable comments and constructive feedback that 
substantially enriched the content and quality of this paper. 
\bibliographystyle{IEEEtran}
\bibliography{citesall}
\begin{IEEEbiography}[{\includegraphics[width=1in,height=1.25in,clip,keepaspectratio]{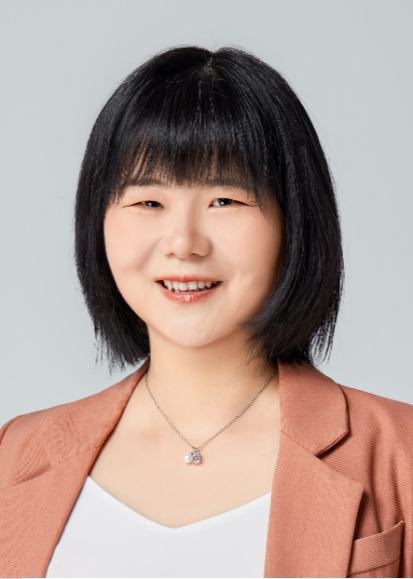}}]{Qiuyun Lyu}
  received the Ph.D. degree from the School of Computer Science and Technology, 
  Hangzhou Dianzi University, China, in 2021. She is currently an associate professor 
  with the School of Cyberspace, Hangzhou Dianzi University. She also works in 
  Key Laboratory of Data Storage and Transmission Technology of Zhejiang Province and 
  Pinghu Digital Technology Innovation Institute Co., Ltd, Hangzhou Dianzi University. 
  Her research interests include data security, self-sovereign identity, 
  and privacy-enhancing technology.
\end{IEEEbiography}
\begin{IEEEbiography}[{\includegraphics[width=1in,height=1.25in,clip,keepaspectratio]{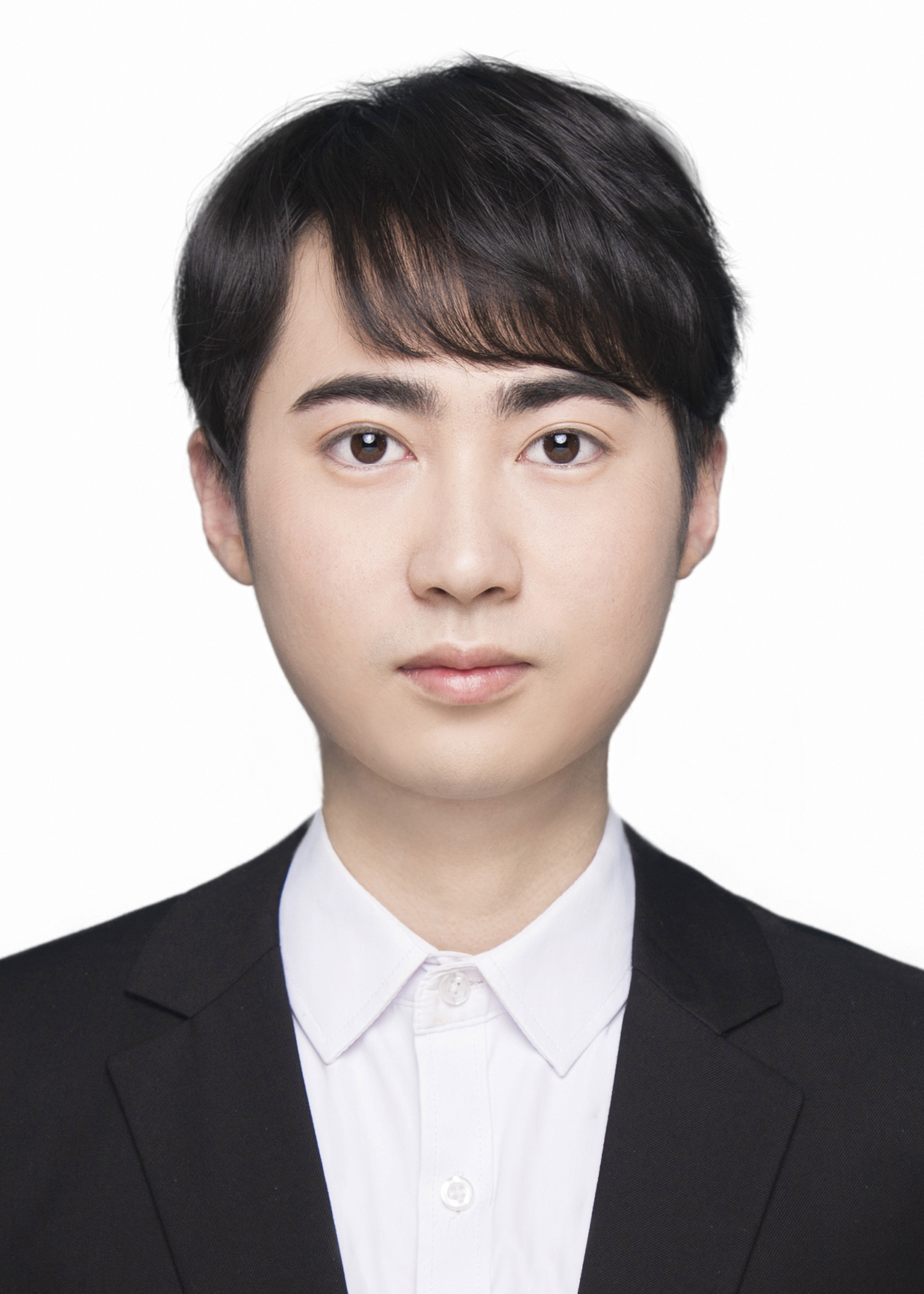}}]{Yilong Zhou}
  received his B.S. degree from the School of Cyberspace, Hangzhou Dianzi University, 
  China, in 2022. He is currently working toward an M.S. degree in Cybersecurity with 
  the School of Cyberspace, Hangzhou Dianzi University. He also works 
  in Pinghu Digital Technology Innovation Institute Co., Ltd, Hangzhou Dianzi University. 
  His research focuses on security and privacy in data sharing, and access control. 
\end{IEEEbiography}
\begin{IEEEbiography}[{\includegraphics[width=1in,height=1.25in,clip,keepaspectratio]{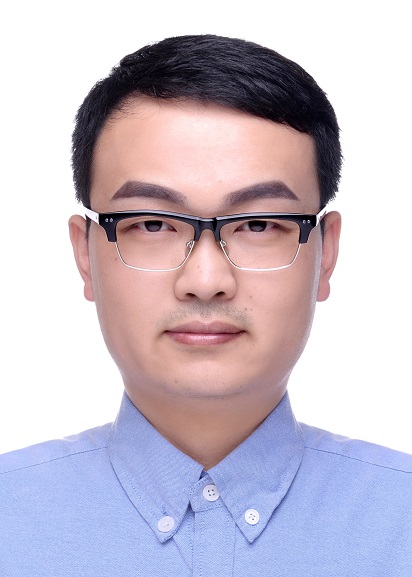}}]{Yizhi Ren}
  received his PhD in Computer software
  and theory from Dalian University of Technology,
  China in 2011. He is currently an professor with
  School of Cyberspace, Hangzhou Dianzi University,
  China. From 2008 to 2010, he was a research
  fellow at Kyushu University, Japan. His current
  research interests include: network security, complex
  network, and trust management. Dr. REN has published over 60 research papers in refereed journals
  and conferences. He won IEEE Trustcom 2018 Best
  Paper Award, CSS2009 Student Paper Award and
  AINA2011 Best Student paper Award. 
\end{IEEEbiography}
\begin{IEEEbiography}[{\includegraphics[width=1in,height=1.25in,clip,keepaspectratio]{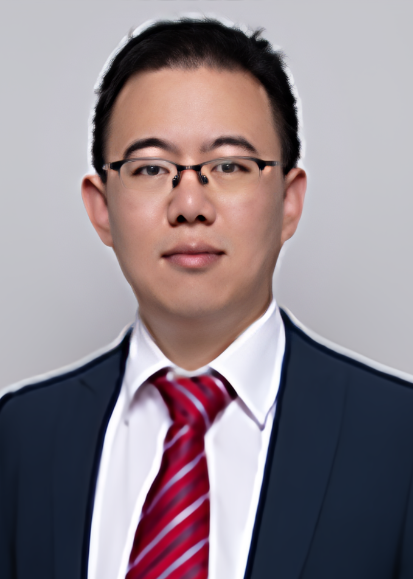}}]{Zheng Wang}
  received the Ph.D. degree in Software Engineering from Dalian 
  University of Technology, China, in 2016. Now He is an associate professor 
  with School of Cyberspace, and vice dean of ZhuoYue Honors College, 
  Hangzhou Dianzi University, China. His research interests include complex 
  networks, network security, and artiﬁcial intelligence. 
\end{IEEEbiography}
\begin{IEEEbiography}[{\includegraphics[width=1in,height=1.25in,clip,keepaspectratio]{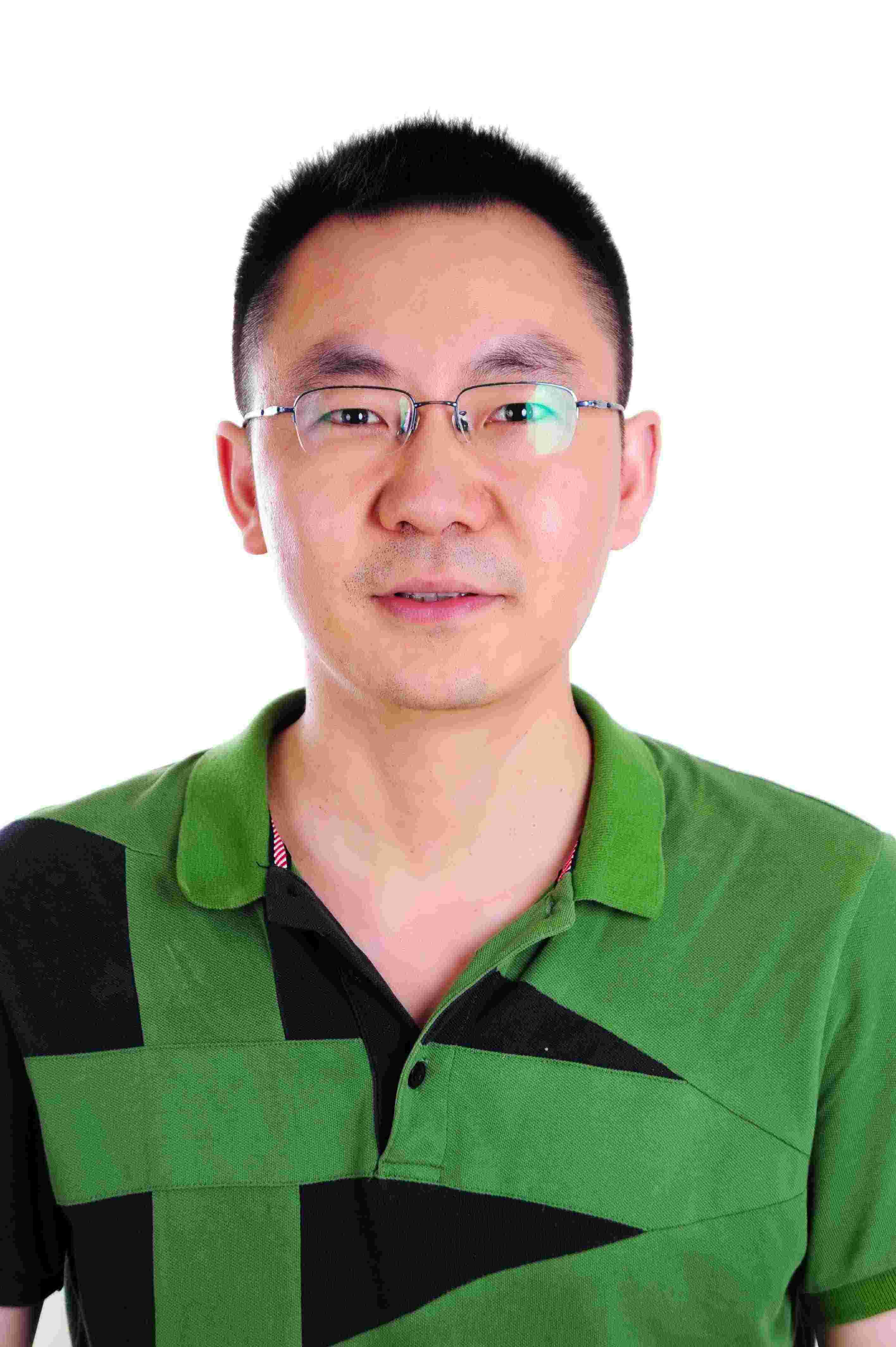}}]{Yunchuan Guo}
  (M'14) received the B.S. and M.S. degrees in computing science and 
  technology, Guilin, China, in 2000 and 2003, respectively, and the Ph.D. 
  degree in computing science and technology from the Institute of Computing 
  Technology, Chinese Academy of Science, Beijing, China, in 2011. 
  He is currently a Professor with the Institute of Information Engineering, 
  Chinese Academy of Sciences. His current research interests include network 
  security and access control. 
\end{IEEEbiography}
\vfill
\end{document}